\documentclass[journal,10pt,twocolumn]{IEEEtran}
\usepackage{amssymb}
\usepackage{algorithm}
\usepackage{algpseudocode}
\usepackage{algorithmicx}
\usepackage{amsfonts}
\usepackage{amsmath}
\usepackage{amssymb}
\usepackage{amsthm}
\usepackage{array}
\usepackage{bbding}
\usepackage{bigints}
\usepackage{booktabs}
\usepackage{cite}
\usepackage{cleveref}
\usepackage{color}
\usepackage{diagbox}
\usepackage{epsfig,latexsym}
\usepackage{epstopdf}
\usepackage{graphicx}
\usepackage{fancyhdr}
\usepackage{float}
\usepackage{flushend}
\usepackage{indentfirst}
\usepackage{lastpage}
\usepackage{makecell}
\usepackage{mathtools}
\usepackage{multirow}
\usepackage{pifont}
\usepackage{psfrag}
\usepackage{setspace}
\usepackage{stfloats}
\usepackage{subfloat}
\usepackage{times}
\usepackage{subfigure}

\newcommand{\argmax}{\arg\,\max}

\begin{document}
\title{Performance Analysis and Approximate Message Passing Detection of Orthogonal Time Sequency Multiplexing Modulation}

\author{{Zeping Sui}, {\em Student Member,~IEEE}, {Shefeng Yan}, {\em Senior Member,~IEEE}, {Hongming~Zhang}, {\em Member,~IEEE}, {Sumei Sun}, {\em Fellow,~IEEE}, {Yonghong Zeng}, {\em Fellow,~IEEE}, {Lie-Liang Yang}, {\em Fellow,~IEEE}, {and Lajos Hanzo}, {\em Life Fellow,~IEEE}
\thanks{This work was supported by the National Science Foundation of China Grants 62192711 and 62001056, the China Scholarship Council under Grant 202004910653, and supported by the National Research Foundation, Singapore and Infocomm Media Development Authority under its Future Communications Research \& Development Programme. (\emph{Corresponding author: Shefeng Yan})}
\thanks{L. Hanzo would like to acknowledge the financial support of the 
Engineering and Physical Sciences Research Council projects 
EP/W016605/1, EP/X01228X/1 and EP/X01228X/1 as well as of the European 
Research Council's Advanced Fellow Grant QuantCom (Grant No. 789028)}
\thanks{Zeping Sui and Shefeng Yan are with the Institute of Acoustics, Chinese Academy of Sciences, Beijing 100190, China; and also with the University of Chinese Academy of Sciences, Beijing 100049, China (e-mail: suizeping@mail.ioa.ac.cn; sfyan@mail.ioa.ac.cn).}
\thanks{Hongming Zhang is with the School of Information and Communication Engineering, Beijing University of Posts and Telecommunications, Beijing 100876, China (email: zhanghm@bupt.edu.cn).}
\thanks{Sumei Sun and Yonghong Zeng are with the Institute for Infocomm Research, Agency for Science, Technology and Research, Singapore 138632 (e-mail: sunsm@i2r.a-star.edu.sg; yhzeng@i2r.a-star.edu.sg).}
\thanks{Lie-Liang Yang and Lajos Hanzo are with the Department of Electronics and Computer Science, University of Southampton, Southampton SO17 1BJ, U.K. (e-mail: lly@ecs.soton.ac.uk; lh@ecs.soton.ac.uk).}
}
\vspace{-5em}
\maketitle

\vspace{-8em}
\begin{abstract}
In orthogonal time sequency multiplexing (OTSM) modulation, the information symbols are conveyed in the delay-sequency domain upon exploiting the inverse Walsh Hadamard transform (IWHT). It has been shown that OTSM is capable of attaining a bit error ratio (BER) similar to that of orthogonal time-frequency space (OTFS) modulation at a lower complexity, since the saving of multiplication operations in the IWHT. Hence we provide its BER performance analysis and characterize its detection complexity. We commence by deriving its generalized input-output relationship and its unconditional pairwise error probability (UPEP). Then, its BER upper bound is derived in closed form under both ideal and imperfect channel estimation conditions, which is shown to be tight at moderate to high signal-to-noise ratios (SNRs). Moreover, a novel approximate message passing (AMP) aided OTSM detection framework is proposed. Specifically, to circumvent the high residual BER of the conventional AMP detector, we proposed a vector AMP-based expectation-maximization (VAMP-EM) detector for performing joint data detection and noise variance estimation. The variance auto-tuning algorithm based on the EM algorithm is designed for the VAMP-EM detector to further improve the convergence performance. The simulation results illustrate that the VAMP-EM detector is capable of striking an attractive BER vs. complexity trade-off than the state-of-the-art schemes as well as providing a better convergence. Finally, we propose AMP and VAMP-EM turbo receivers for low-density parity-check (LDPC)-coded OTSM systems. It is demonstrated that our proposed VAMP-EM turbo receiver is capable of providing both BER and convergence performance improvements over the conventional AMP solution.
\end{abstract}
\vspace{-1em}
\IEEEpeerreviewmaketitle

\section{Introduction}\label{Section1}

Next-generation wireless communication systems are expected to communicate reliably via high-mobility channels even in the face of severe inter-carrier interference (ICI) and inter-symbol interference (ISI) \cite{steele1999mobile,hanzo2005ofdm}. With this in mind, orthogonal time-frequency space (OTFS) modulation has been proposed as an attractive candidate for next-generation wireless communications \cite{DBLP:journals/corr/abs-1802-02623,8686339,9610105,8516353}, where the information bits are modulated in the delay-Doppler (DD) domain. Then the two-dimensional orthogonal basis functions of inverse fast Fourier transform (IFFT) and FFT are employed along in the delay and Doppler domains, respectively, for spreading the DD domain symbols across the whole time-frequency (TF) domain. The most appealing aspect of OTFS is that it leverages the DD domain representation of the doubly-selective channel, where the dimension of the channel is reduced to the number of channel coefficients, implying that OTFS exhibits the sparsest channel representation \cite{8424569,9590508}. Moreover, the double-selective time-frequency (TF) channel is converted into the quasi-static DD domain channel based on the two-dimensional orthogonal basis functions, making OTFS capable of handling doubly-selective channels \cite{9508932,li2021performance}. However, the complexity of the OTFS system escalates, since the DD domain symbols are followed by the cascaded inverse symplectic finite Fourier transform (ISFFT)/SFFT and Heisenberg/Wigner transform pairs, especially for high numbers of time slots and subcarriers \cite{yuan2021iterative,9536449}.

As a relative of OTFS, orthogonal time sequency multiplexing (OTSM) modulation was first proposed in \cite{thaj2021orthogonal1,thaj2021orthogonal}, where sequency is defined as the number of the signal sign changes per second. In contrast to OTFS, the information symbols are arranged in the delay-sequency (DS) domain, and the delay-time (DT) domain frame is then obtained after invoking the inverse Walsh-Hadamard transform (IWHT) in the sequency domain in OTSM systems. The input-output relationship of OTSM proposed in \cite{thaj2021orthogonal} is derived based on the Fourier relationship between a pair of functions within the Bello system function family, i.e. the delay-spread function and the delay-Doppler-spread function \cite{steele1999mobile}. Therefore, OTSM is also capable of exploiting the sparse representation of the time-varying multipath channel in the delay-sequency domain in a fashion reminiscent of the delay-Doppler domain channel representation of OTFS. Furthermore, OTSM may also be viewed as a single-carrier modulation scheme, where the ISI introduced by the delay spread and Doppler spread can be processed separately in delay and sequency domains, respectively \cite{thaj2021orthogonal}. Additionally, zero-padding (ZP) is integrated into OTSM systems to alleviate inter-block interference (IBI). As a benefit, OTSM is capable of attaining a similar bit error ratio (BER) performance to OTFS at a lower complexity in quasi-static and high-mobility channels, since the IWHT only needs addition and subtraction operations, rather than full-known multiplications.
\begin{table*}[t]
\footnotesize
\begin{center}
\caption{Contrasting our contributions to the state-of-the-art}
\label{table1}
\begin{tabular}{l|c|c|c|c|c|c|c}
\hline
Contributions & \textbf{Our paper} & \cite{9610105} & \cite{8424569} & \cite{li2021performance} & \cite{yuan2021iterative} & \cite{thaj2021orthogonal1} & \cite{thaj2021orthogonal} \\
\hline
\hline
OTSM & \checkmark &  &  &  &  & \checkmark & \checkmark \\  
\hline
BER performance analysis of OTSM & \checkmark &  &  &  &  &  & \\  
\hline
VAMP-based detection & \checkmark &   &  &  &   &  & \\  
\hline Statistics of transmitted symbols & \checkmark & \checkmark & \checkmark &  & \checkmark &  & \\  
\hline
EM noise variance estimation & \checkmark &  &  & &  &  & \\ 
\hline
Auto-tuning & \checkmark &  &  &  &  &  & \\ 
\hline
Input-output relationship of generalized OTSM & \checkmark &  &  &  &  &  & \\ 
\hline
Coded system & \checkmark &  &  & \checkmark & \checkmark &  &  \checkmark \\ 
\hline
Turbo receiver & \checkmark &  &  &  & \checkmark &  &  \checkmark \\ 
\hline
Channel estimation imperfection & \checkmark &  &  &  &  &  &  \checkmark \\
\hline
\end{tabular}
\end{center}
\vspace{-6mm}
\end{table*}

Nonetheless, there are several open problems in the context of OTSM that require further investigation. Firstly, although the system model and the BER performances of several detectors designed for ZP-OTSM have been illustrated in \cite{thaj2021orthogonal},  the generalized input-output relationship of OTSM dispensing with ZP and its formal BER analysis have not been unveiled. Furthermore, since the information symbols are mapped to the two-dimensional DS-domain grids, it is challenging to design low-complexity detectors having good performance. In \cite{thaj2021orthogonal1}, both a single-tap time domain (TD) and an iterative detector based on the Gauss-Seidel (GS) method have been proposed. However, the BER performance of the single tap detector degrades in high-mobility channels, while the GS detector includes matrix inversion at the cost of high complexity. It should be noted that none of the above detection algorithms exploits the sparsity of the DS-domain channel matrix, and neither do they exploit the statistics of the modulation schemes. We will fill this knowledge-gap and design efficient detectors based on approximate message passing (AMP) algorithms. Let us now review some of the existing works below to gain insight into designing AMP-based OTSM detectors.

The AMP algorithm that relies on loopy belief propagation (BP) relying on beneficial approximations has been proposed as a low-complexity signal recovery algorithm in \cite{donoho2009message}. The so-called ``Onsager term'' in the expressions of the AMP algorithm leads to a better sparsity-undersampling tradeoff than that of the iterative thresholding algorithms \cite{9507331}. Nonetheless, the AMP algorithm was initially conceived for zero-mean independent and identically distributed (i.i.d.) Gaussian measurement matrices. For a general sensing matrix, the performance of the AMP algorithm would suffer significantly. In \cite{rangan2019vector}, Rangan \emph{et al.} proposed the vector AMP (VAMP) algorithm based on non-loopy factor graphs, which is suitable for right-orthogonally invariant matrices. Therefore, VAMP can handle a much broader class of matrices and achieves a significantly better mean square error (MSE) performance at a similar complexity to AMP. Given these compelling properties, the VAMP algorithm has been invoked for applications such as learning unknown hyper-parameters \cite{DBLP:journals/corr/FletcherSSR17} and bilinear recovery \cite{8712432}. However, the above-mentioned design philosophies have not been applied to the OTSM scheme.

Our novel contributions are boldly and explicitly contrasted to the existing literature in Table \ref{table1} at a glance, which are further detailed as follows:
\begin{itemize}
\item We first conceive the input-output relationship of generalized OTSM systems dispensing with ZP, which are different from the systems in~\cite{thaj2021orthogonal,thaj2021orthogonal1}. Then, the conditional pairwise error probability (PEP) of OTSM systems for a given DS-domain input-output relationship is derived by investigating the modified Euclidean distance of pairwise error events, and the unconditional PEP (UPEP) is further conceived based on PEP. Moreover, based on the UPEP and the union bounding technique, we derive a tight BER upper bound of OTSM systems in closed form, where the channel estimation imperfection is also considered.
	\item We propose a low-complexity VAMP-aided OTSM detector based on the \emph{a priori} information of the modulation scheme, namely the VAMP-based expectation-maximization (VAMP-EM) detector. We first investigate the conventional AMP-aided detector. Then, based on the statistics of the transmitted symbols, the VAMP-aided detector is proposed by relying on the vector-valued factor graph. Thirdly, we exploit the EM algorithm to learn the noise variance to be incorporated in the VAMP detector. Hence, the proposed VAMP-EM detector provides more precise symbol vector estimates under unknown noise variance scenarios than the conventional AMP detector. As a further advance, an auto-tuning strategy is conceived for estimating the parameters $\gamma_1$ and $\gamma_2$ at every iteration in order to improve the efficiency of VAMP-EM. Simulation results are provided for characterizing the overall performance of the VAMP-EM detector. It is demonstrated that the proposed VAMP-EM achieves a significant BER and convergence rate improvement. Furthermore, the VAMP-EM detector is capable of striking an attractive BER vs. complexity trade-off in uncoded OTSM systems.
	\item We extend the AMP and VAMP-EM detectors to turbo receivers designed for the low-density parity-check (LDPC) coded OTSM systems. Specifically, the attainable performances of the AMP and VAMP-EM turbo detectors conceived are evaluated at different LDPC-coded rates and different numbers of iterations, and the powerful semi-analytical tools of extrinsic information transfer (EXIT) charts are invoked to visually characterize the soft-information flow between the decoders and our proposed detectors. Our simulation results demonstrate that the proposed VAMP-EM turbo receiver attains both a considerable BER performance gain and a convergence speed-up.
\end{itemize}

The rest of the paper is organized as follows. The system model is portrayed in Section \ref{Section2}, followed by our error performance analysis in Section \ref{Section3}. In Section \ref{Section4}, the proposed AMP-aided detection algorithms are discussed, while the AMP and VAMP-EM-based turbo receivers are introduced in Section \ref{Section5}. Our simulation results are illustrated in Section \ref{Section6}. Finally, our conclusions are offered in Section \ref{Section7}.

\emph{Notation:} We use the following notation throughout this paper. $\mathbb{B}$ denotes the bit set consisting of $\left\{0,1\right\}$; $\mathbb{E}[\cdot]$, $\text{tr}\left\{\cdot\right\}$, $[\cdot]_{N}$ and $\Re\left\{\cdot\right\}$ represent the expectation, trace, module-$N$ and real part operators respectively; Matrices and vectors are denoted by upper- and lower-case boldface letters, respectively; For a length-$n$ vector $\pmb{u}$, the empirical averaging operator can be given as $\left\langle\pmb{u}\right\rangle=\frac{1}{N}\sum_{n=1}^N u(n)$; $\text{vec}(\pmb{A})$ denotes the vector formulated by stacking the columns of $\pmb{A}$ to obtain a single column vector matrix, and $\text{vec}^{-1}(\pmb{a})$ denotes the inverse vectorization operation to form the original matrix; $\otimes$ is the Kronecker product of two matrices; $I_0(\cdot)$ the zero-order modified Bessel function of the first kind; $\mathcal{CN}(\pmb{a},\pmb{B})$ denotes the complex Gaussian distribution with mean vector $\pmb{a}$ and covariance matrix $\pmb{B}$; For a matrix $\pmb{A}$, $(\pmb{A})^T$, $(\pmb{A})^{\ast}$, $(\pmb{A})^H$, $(\pmb{A})^{-1}$ and $|\pmb{A}|^2$ represent its transpose, conjugate, conjugate transpose, inverse and element-wise modulo squared operation respectively; $\pmb{\mathcal{W}}_{N}$ stands for the $N$-point IWHT matrix; $\pmb{I}_N$ and $\pmb{e}_N(n)$ denote the $N$-dimensional identity matrix and its $n$th column; $\pmb{0}_{u\times v}$ is the $u\times v$ zero matrix; $Q(x)=\int_x^{+\infty}\frac{1}{\sqrt{2\pi}}\exp\left(-\frac{1}{2}t^2\right)dt$ and $\delta(\cdot)$ represent the $Q$-function and the delta function; $\pmb{x}\cdot\pmb{y}$ and $\pmb{x}\cdot/\pmb{y}$ denotes the element-wise product and division between vectors $\pmb{x}$ and $\pmb{y}$; $\pmb{1}$ and $\pmb{0}$ denote the all-ones and all-zeros vectors with a proper length, respectively; $H(q)\triangleq-\int q(x)\ln q(x)\mathrm{d}x$ denotes the differential entropy of $x\sim q$; $D_\text{KL}\triangleq \int q(x)\ln\frac{q(x)}{p(x)}\mathrm{d}x$ is the \emph{Kullback-Leibler divergence} from $p$ to $q$; $\mathcal{U}[a,b]$ represents the uniform distribution in the interval $[a,b]$.
\section{System Model}\label{Section2}
\subsection{Transmitter Description}
\begin{figure*}[htbp]
\centering
\includegraphics[width=0.8\linewidth]{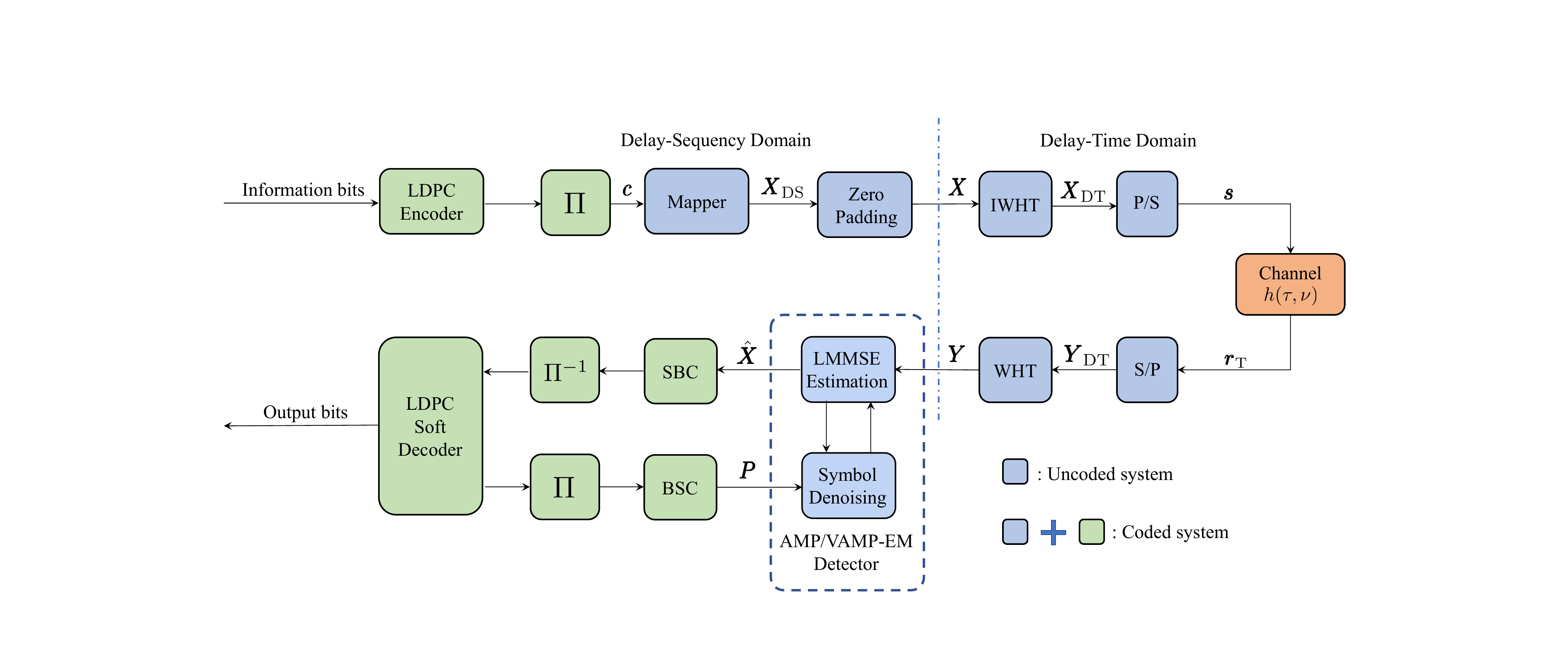}
\caption{Illustration of the uncoded and LDPC coded OTSM systems, where $\Pi$ and $\Pi^{-1}$ denote the turbo interleaver and deinterleaver, respectively.}
\label{Figure1}
\vspace{-1.5em}
\end{figure*}
\begin{figure*}[htbp]
\centering
\includegraphics[width=0.8\linewidth]{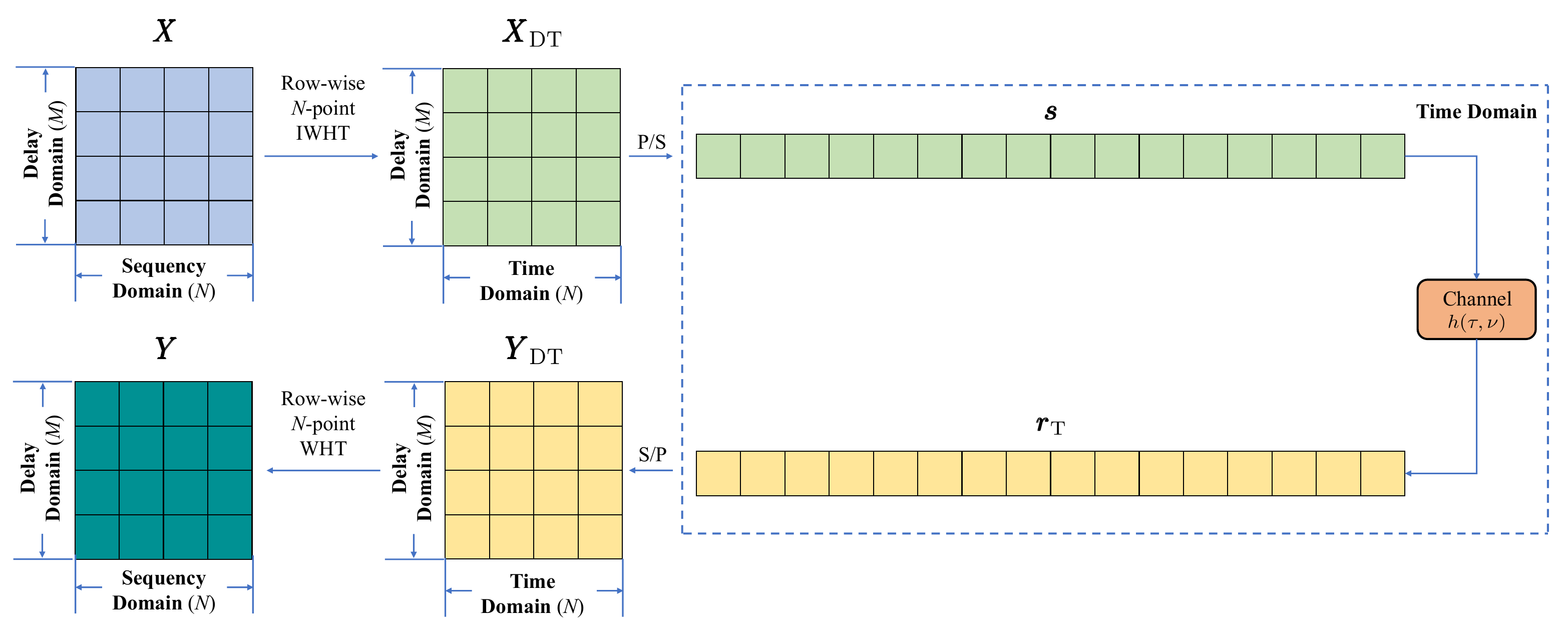}
\caption{Relationship between the grids in the DS-domain, DT-domain, and time domain of the OTSM systems with $M=4$ and $N=4$.}
\label{Figure2}
\vspace{-1em}
\end{figure*}
We consider a ZP-OTSM system having a bandwidth $B=M\Delta f$ and the frame duration $T_f=NT$, where $T$ represents the symbol duration and $\Delta f$ is the sampling frequency interval, while $M$ and $N$ denote the maximum value of delay and sequency indices. Moreover, the system is critically sampled, i.e., we have $T\Delta f=1$. Throughout this section, we denote the number of data subcarriers by $L_\text{D}=M-L_\text{ZP}$, where $L_\text{ZP}$ represents the length of ZP. The OTSM system is shown in Fig.~\ref{Figure1}. At the transmitter, an i.i.d information bit sequence $\pmb{b}\in\mathbb{B}^{RL_b}$ is encoded by a rate-$R$ LDPC encoder. Hence the LDPC-coded bit sequence $\pmb{c}=[\pmb{c}_0,\ldots,\pmb{c}_{J-1}]\in\mathbb{B}^{L_b}$ is obtained after interleaving, where $\pmb{c}_{j}=[c_{j}(1),\ldots,c_{j}(\log_2 K)]$ for $j=0,\ldots,J-1$ and $L_\text{b}=NL_\text{D}\log_2 K=J\log_2 K$. Here $K$ is the modulation order and the normalized $K$-ary alphabet is denoted $\mathcal{A}=\{a_1,\ldots,a_K\}$, with each constellation point $a_k$ corresponds to a bit pattern $\pmb{s}_k=[s_k(1),\ldots,s_k(\log_2 K)]$ for $k=1,\ldots,K$. Then the coded bit sequence is mapped onto $J$ quadrature amplitude modulation (QAM) symbols that can be formulated as $\pmb{x}_\text{D}=[\pmb{x}^T_0,\ldots,\pmb{x}^T_{L_\text{D}-1}]^T\in\mathbb{C}^{NL_\text{D}\times1}$, where $\pmb{x}_{l_d}=[x_{l_d}(0),x_{l_d}(1),\ldots,x_{l_d}(N-1)]^T\in\mathbb{C}^{N\times 1}$ for $l_d=0,\ldots,L_\text{D}-1$, and we have $x_{l_d}(n)\in\mathcal{A}, \forall l_d,n$. The DS-domain data matrix $\pmb{X}_\text{DS}$ can be obtained by arranging symbol vectors on the DS-domain lattice $\Gamma=\left\{\left(\frac{l}{M\Delta f},\frac{n}{NT}\right),l=0,\ldots,L_\text{D}-1,n=0,\ldots,N-1\right\}$ as $\pmb{X}_\text{DS}=[\pmb{x}_0,\ldots,\pmb{x}_{L_\text{D}-1}]^T\in\mathbb{C}^{L_D\times N}$. The relationship between the sampling grids in the DS-domain, the DT-domain, and the TD is demonstrated in Fig. \ref{Figure2}, which is detailed below. Firstly, we obtain the transmitted DS-domain matrix after inserting the ZP, which can be written as $\pmb{X}=\pmb{\Upsilon}_\text{ZP}\pmb{X}_\text{DS}\in\mathbb{C}^{M\times N}$, where $\pmb{\Upsilon}_\text{ZP}=[\pmb{I}_{L_\text{D}},\pmb{0}_{{L_\text{D}}\times L_\text{ZP}}]^T$ is the $(M\times L_D)$-element mapping matrix. Then IWHT is applied to every row of $\pmb{X}$ to obtain the DT-domain transmitted matrix, yielding $\pmb{X}_\text{DT}=\pmb{X}\pmb{\mathcal{W}}_N$. The elements of the $(N\times M)$-dimensional WHT matrix $\pmb{\mathcal{W}}_N$ are given as $\mathcal{W}_N(n,m)={\tilde{W}(n,m/N+0.5/N)}/{\sqrt{N}}$, where $\tilde{W}(n,\lambda)$ represents the continuous Walsh functions within $0\leq\lambda\leq 1$ \cite{thaj2021orthogonal}. Hence, the TD transmitted signal can be attained as $\pmb{s}=\text{vec}(\pmb{X}_\text{DT})=\pmb{P}(\pmb{I}_M\otimes\pmb{\mathcal{W}}_N)\pmb{x}$, where $\pmb{x}=[\pmb{x}_\text{D}^T,\pmb{0}_{1\times L_\text{ZP}}]^T\in\mathbb{C}^{MN\times1}$ and the $\pmb{P}$ is the $(MN\times MN)$-dimensional row-column permutation matrix, which is also known as the perfect shuffle matrix \cite{van2000ubiquitous}, formulated as:
\begin{align}
	\pmb{P}=
	\begin{bmatrix}
		\pmb{I}_N\otimes\pmb{e}_M^T(0)\\
		\pmb{I}_N\otimes\pmb{e}_M^T(1)\\	
		\vdots\\
		\pmb{I}_N\otimes\pmb{e}_M^T(M-1)\\		
		\end{bmatrix}.
\end{align}
Based on the fact that $\pmb{P}^T=\pmb{P}^{-1}$ and $\pmb{A}\otimes\pmb{B}=\pmb{P}(\pmb{B}\otimes\pmb{A})\pmb{P}^T$ for square matrices $\pmb{A}$ and $\pmb{B}$, we have
\begin{align}\label{Eq5}
	\pmb{s}=(\pmb{\mathcal{W}}_N\otimes\pmb{I}_M)(\pmb{P}\pmb{x}).
\end{align}
\subsection{Channel Model}
Overall a time-varying multipath channel with $P$ taps is considered, whose channel impulse response (CIR) can be given in the delay-Doppler-spread Bello function form as \cite{steele1999mobile} :
\begin{align}\label{Eq6}
h(\tau,\nu)=\sum_{i=1}^{P}h_{i}\delta(\tau-\tau_i)\delta(\nu-\nu_i),
\end{align}
where $h_i$, $\tau_i$ and $\nu_i$ denote the complex-valued path gain, symbol duration-normalized delay and normalized Doppler shift corresponding to the $i$th path, respectively. Here, we assume that $h_i\sim\mathcal{CN}(0,1/P),\forall i$, which is independent of $\tau_i$ and $\nu_i$ \cite{yuan2021iterative}. The sampling values of delay shifts and Doppler shifts are denoted as $l_i=a_i+\alpha_i$ and $k_i=b_i+\beta_i$, respectively, where $a_i$ and $b_i$ represent the integer-valued delay and Doppler indices, while $\alpha_i,\beta_i\in\mathcal{U}[-\frac{1}{2},\frac{1}{2}]$ are the non-integer fractional components. Therefore, the normalized delay and Doppler shifts are defined as $\tau_i=\frac{l_i}{M\Delta f},\quad\nu_i=\frac{k_i}{NT}$. The maximum delay of the channel is $\tau_\text{max}=l_\text{max}/M\Delta f$ where $l_\text{max}=\max\{\mathcal{L}\}=\max\{l_1,\ldots,l_P\}$. Furthermore, we set $L_\text{ZP}=l_\text{max}+1$ for mitigating the IBI between neighboring data blocks. Based on the intrinsic Fourier relationships among the Bello functions, the continuous time-varying CIR can be expressed with the aid of the input delay-spread function as \cite{steele1999mobile}
\begin{align}\label{Eq8}
	h(\tau,t)=\int h(\tau,\nu)e^{j2\pi\nu(t-\tau)}d\nu=\sum_{i=1}^P h_i e^{j2\pi\nu_i(t-\tau_i)},
\end{align}
which is processed by sampling at the grid points $\{\tau=\frac{l}{M \Delta f},t=\frac{q}{M \Delta f}\}$, yielding
\begin{align}\label{Eq9}
	h(l,q)= \sum_{i=1}^P h_i z_i^{(q-l)}\delta(l-l_i),
\end{align}
where $z_i=e^{j2\pi\frac{k_i}{MN}}$ and $q=0,1,\ldots,MN-1$.

\subsection{Received Signals}
\vspace{-0.5em}
Given the delay-Doppler-spread function $h(\tau,\nu)$, the received TD signal can be written as \cite{steele1999mobile}
\begin{align}\label{Eq10}
	r_\text{T}(t)=\int\int h(\tau,\nu)s(t-\tau)e^{j2\pi\nu(t-\tau)}d\tau d\nu+n(t),
\end{align}
where $n(t)$ is the complex additive white Gaussian noise (AWGN) with zero mean and variance $N_0=1/\gamma_n$. Here, $\gamma_n$ denotes the inverse noise variance, i.e., the noise precision. Therefore, by sampling the received signal at $\{t=\frac{q}{M \Delta f},q=0,1,\ldots,MN-1\}$, we can formulate the corresponding discrete-time received signals based on \eqref{Eq9} and \eqref{Eq10} as
\begin{align}\label{Eq11}
	r_\text{T}(q)=\sum_{l\in\mathcal{L}}h(l,q)s(q-l)+n(q),
\end{align}
\begin{figure*}[t]
\centering
\subfigure[]{\label{Figure3-1}\includegraphics[width=0.32\linewidth]{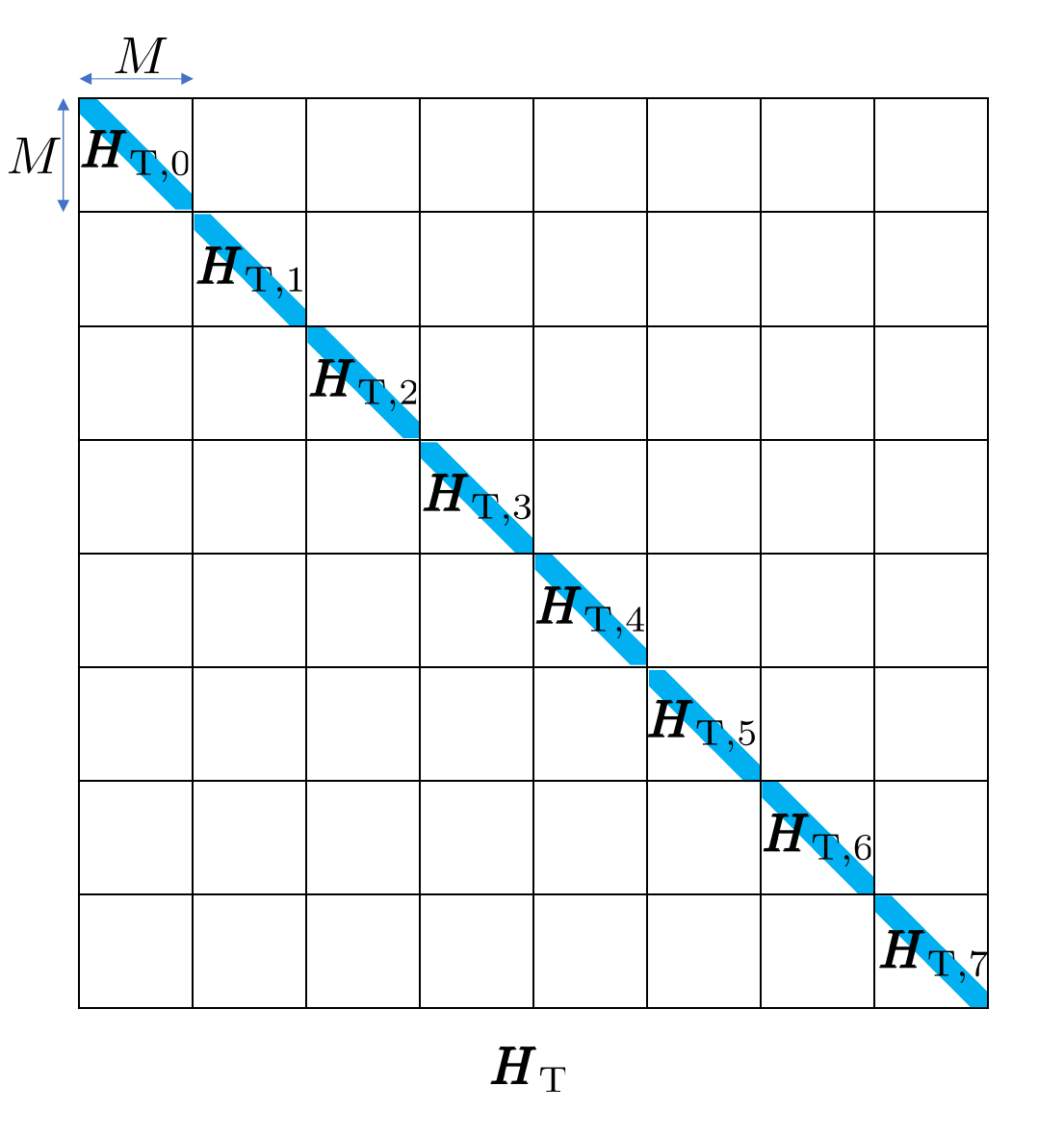}}
\subfigure[]{\label{Figure3-2}\includegraphics[width=0.32\linewidth]{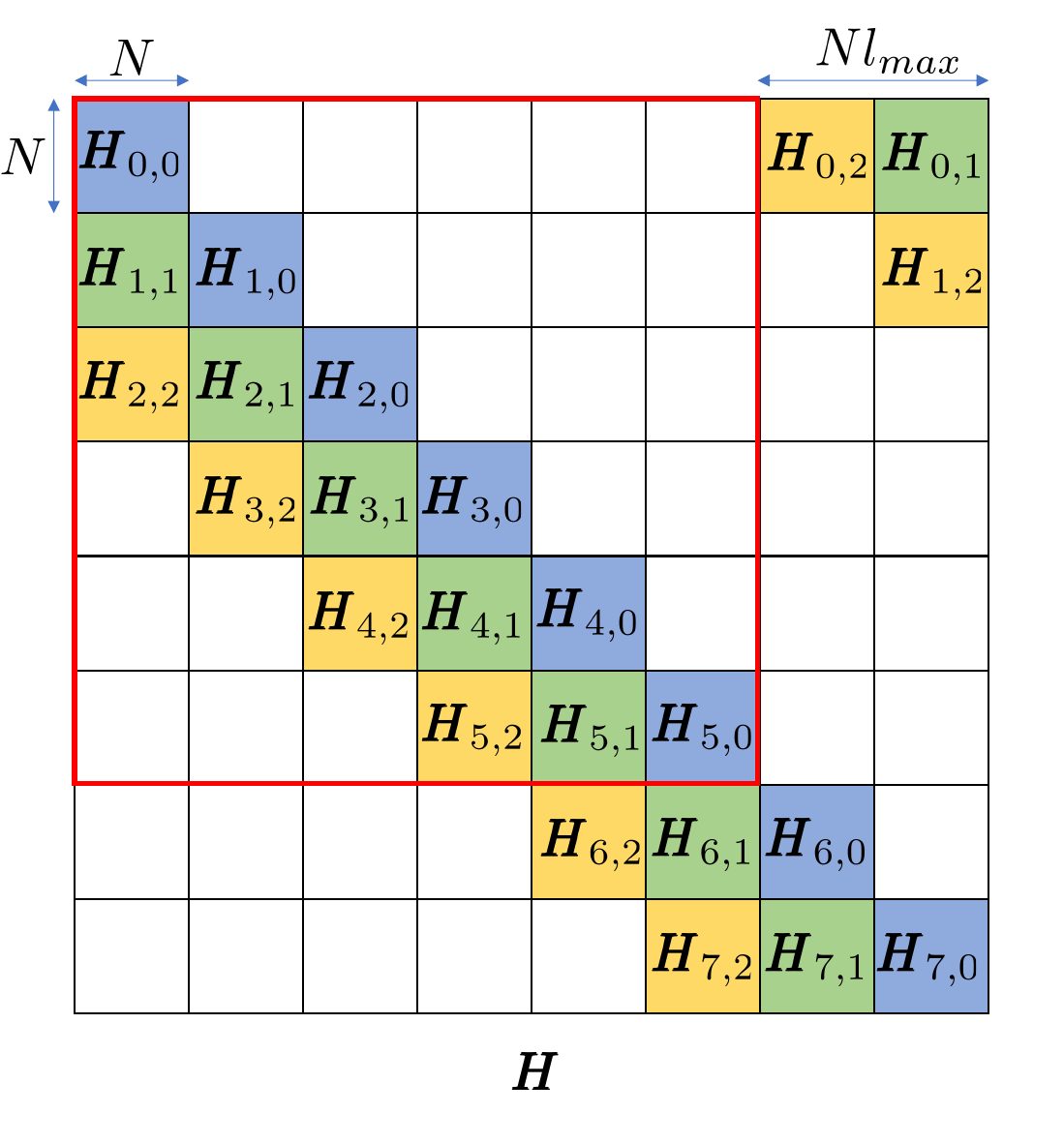}}
\caption{Illustration of the effective time-domain channel matrix $\pmb{H}_\text{T}$ and delay-sequency domain channel matrix $\pmb{H}$ with $M=N=8$ and $l_\text{max}=2$.}
\vspace{-5mm}
\label{Figure3}
\end{figure*}
where $s(q)=s(t)|_{t=\frac{q}{M\Delta f}}$ and $n(q)=n(t)|_{t=\frac{q}{M\Delta f}}$. Therefore, the $(MN)$-samples of the TD received signal and noise can be written as $\pmb{r}_\text{T}=[r_\text{T}(0),\ldots,r_\text{T}(MN-1)]^T$ and $\pmb{n}=[n(0),\ldots,n(MN-1)]^T$, respectively. Hence, based on the TD input-output relationship of \eqref{Eq10}, the corresponding received symbol vector is given by
\begin{align}\label{Eq14}
	\pmb{r}_\text{T}=\pmb{H}_\text{T} \pmb{s}+\pmb{n},
\end{align}
where $\pmb{H}_\text{T}=\text{diag}[\pmb{H}_{\text{T},0},\ldots,\pmb{H}_{\text{T},N-1}]\in\mathbb{C}^{MN\times MN}$ is the effective TD channel matrix, and $\pmb{H}_{\text{T},n}\in\mathbb{C}^{M\times M}$ is the $n$th block channel matrix for $n=0,\ldots,N-1$. The non-zero valued elements of $\pmb{H}_\text{T}$ can be expressed as $H_\text{T}(q,q-l)=h(l,q)$. By rewriting all the elements of $\pmb{H}_\text{T}$ as $H(a,b)$, it can be readily shown that $l_{\text{max}}=\max\{a-b\}$, $\forall a,b$. Therefore, as shown in Fig. \ref{Figure3} \subref{Figure3-1}, the bandwidth of $\pmb{H}_\text{T}$ is $l_\text{max}$ and there are $L\leq l_\text{max}+1$ non-zero values in each row and column of $\pmb{H}_\text{T}$  \cite{thaj2021orthogonal}.
After the inverse vectorization process, the received DT-domain symbol matrix can be formulated as $\pmb{Y}_\text{DT}=\text{vec}^{-1}(\pmb{r}_\text{T})=[\pmb{y}_{\text{DT},0},\ldots,\pmb{y}_{\text{DT},M-1}]^T$.
The received DS-domain matrix $\pmb{Y}$ is obtained by applying the WHT to each row of $\pmb{Y}_\text{DT}$, yielding $\pmb{Y}=\pmb{Y}_\text{DT}\pmb{\mathcal{W}}_N=[\pmb{y}_0,\ldots,\pmb{y}_{M-1}]^T$. Hence, the received DS-domain symbol vector is given by
\begin{align}\label{Eq17}
	\pmb{y}=(\pmb{I}_M\otimes \pmb{\mathcal{W}}_N)(\pmb{P}^T\pmb{r}_\text{T}),
\end{align}
where $\pmb{y}=[\pmb{y}_0^T,\ldots,\pmb{y}_{M-1}^T]^T$.
Finally, based on \eqref{Eq5}, \eqref{Eq14} and \eqref{Eq17}, the DS-domain input-output relationship can be expressed as
\begin{align}\label{Eq18}
	\pmb{y}=\pmb{H}\pmb{x}+\bar{\pmb{n}},
\end{align}
where the DS-domain channel matrix and the equivalent AWGN vector can be respectively formulated as $\pmb{H}=(\pmb{I}_M\otimes \pmb{\mathcal{W}}_N)(\pmb{P}^T\pmb{H}_\text{T} \pmb{P})(\pmb{I}_M\otimes \pmb{\mathcal{W}}_N)$ and $\bar{\pmb{n}}=(\pmb{I}_M\otimes \pmb{\mathcal{W}}_N)(\pmb{P}^T\pmb{n})$. Since the transform matrix $(\pmb{I}_M\otimes \pmb{\mathcal{W}}_N)\pmb{P}^T$ is unitary, we can readily show that $\bar{n}(q)\sim\mathcal{CN}(0,\gamma_n^{-1})$, $\forall q$. Therefore, the average SNR per symbol can be expressed as $\gamma_s=\gamma_n$. As illustrated in Fig. \ref{Figure3} \subref{Figure3-2}, the strictly upper triangular sub-matrices of $\pmb{H}$ can be ignored since ZP is used, while the bandwidth of $\pmb{H}$ is $Nl_\text{max}$, and each row and column of $\pmb{H}$ have $NL$ non-zero elements. The sub-matrices in $\pmb{H}$ can be written as $\pmb{H}_{m,l}=\pmb{\mathcal{W}}_N\bar{\pmb{H}}_{m,l}\pmb{\mathcal{W}}_N$ \cite{thaj2021orthogonal}, where $\bar{\pmb{H}}_{m,l}=\text{diag}[\bar{h}_{m,l}(0),\ldots,\bar{h}_{m,l}(N-1)]$ and $\bar{h}_{m,l}=h(l,m+nM)$ for $m=0,\ldots,M-1$ and $n=0,\ldots,N-1$.

\section{Error Performance Analysis}\label{Section3}
In this section, we investigate the BER performance of generalized OTSM systems dispensing with ZP, which differs from the ZP-OTSM systems in Section \ref{Section2}. Throughout this section, we assume that perfect channel state information (CSI) is available at the receiver, and the optimum ML detector detailed in Section \ref{Section4-1} is employed. For notational consistency, we also use $\left\{\pmb{s},\pmb{r}_\text{T},\pmb{x},\pmb{y}\right\}$ to denote the TD and DS-domains transmitted and received symbol vectors, respectively. We commence by detailing the input-output relationship of the generalized OTSM systems. Then the conditional PEP (CPEP) and the UPEP of the OTSM system are derived. Finally, the asymptotically tight upper bound on the BER of generalized OTSM systems is derived.

To alleviate the ISI between OTSM frames, only a cyclic prefix (CP) length of $L_\text{CP}=l_\text{max}+1$ is appended to the transmitted signal. The TD input-output relationship given by \eqref{Eq11} can be further formulated as
\begin{align}\label{Eq22}
	r_\text{T}(q)=\sum_{i=1}^{P} h_i e^{j2\pi \frac{k_i(q-l_i)}{MN}}s(\left[q-l_i\right]_{MN})+n(q).
\end{align}
where $q=0,1,\ldots,MN-1$.
 Therefore, \eqref{Eq22} can be written in a vectorial form as
\begin{align}\label{Eq23}
	\pmb{r}_\text{T}=\pmb{G}\pmb{s}+\pmb{n},
\end{align}
Specifically, the TD channel matrix can be derived as $\pmb{G}=\sum_{i=1}^P h_i \pmb{\Pi}^{l_i}\pmb{\Delta}^{k_i}$, where the permutation matrix $\pmb{\Pi}$ can be obtained by employing a cyclic forward shift to the rows of $\pmb{I}_{MN}$ \cite{9590508}, and $\pmb{\Delta}=\text{diag}[z^0,\ldots,z^{MN-1}]$ with $z=e^{j\frac{2\pi}{MN}}$.
Given the transformation relationships shown in Fig. \ref{Figure2}, by substituting \eqref{Eq5} and \eqref{Eq17} into \eqref{Eq23}, the DS-domain input-output relationship can be derived as
\begin{align}\label{Eq26}
	\pmb{y}=\bar{\pmb{H}}\pmb{x}+\bar{\pmb{n}},
\end{align}
where $\pmb{x}\in \mathcal{A}^{NM\times1}$. Specifically, the DS-domain channel matrix can be written as
\begin{align}\label{Eq27}
	\bar{\pmb{H}}=\sum_{i=1}^P h_i (\pmb{I}_M\otimes \pmb{\mathcal{W}}_N)(\pmb{P}^T\pmb{\Pi}^{l_i}\pmb{\Delta}^{k_i}\pmb{P})(\pmb{I}_M\otimes \pmb{\mathcal{W}}_N).
	\end{align}
\vspace{-1em}
\subsection{Ideal Channel Conditions}
Substituting \eqref{Eq27} into \eqref{Eq26}, it can be shown that \eqref{Eq26} can be further formulated as
\begin{align}\label{Eq28}
	\pmb{y}=\pmb{\Phi}(\pmb{x})\pmb{h}+\pmb{\bar{n}},
	\end{align}
where $\pmb{h}=[h_1,\ldots,h_P]^T\in\mathbb{C}^{P\times 1}$ is the channel coefficient vector. The concatenated equivalent codeword matrix $\pmb{\Phi}(\pmb{x})\in\mathbb{C}^{MN\times P}$ can be expressed as
\begin{align}\label{Eq29}
	\pmb{\Phi}(\pmb{x})=[\underbrace{\pmb{D}_1\pmb{x}}_{MN\times 1}\quad\pmb{D}_2\pmb{x}\quad \ldots\quad\pmb{D}_P \pmb{x}],
	\end{align}
where $\pmb{D}_p$ is given by $\pmb{D}_p=(\pmb{I}_M\otimes \pmb{\mathcal{W}}_N)(\pmb{P}^T\pmb{\Pi}^{l_i}\pmb{\Delta}^{k_i}\pmb{P})(\pmb{I}_M\otimes \pmb{\mathcal{W}}_N)$, for $p=1,\ldots,P$.

Let us consider the pairwise error event $\{\pmb{x}^c\rightarrow\pmb{x}^e\}$, where $\pmb{x}^c$ is the transmitted symbol vector, and $\pmb{x}^e$ is the error-infested received symbol vector. We define the error vector as $\pmb{e}=\pmb{x}^c-\pmb{x}^e$. Hence, the modified Euclidean distance under this event can be expressed as
\begin{align}\label{Eq31}
\delta=||\pmb{\Phi}(\pmb{e})\pmb{h}||_2^2=\pmb{h}^H\pmb{\Phi}(\pmb{e})^H\pmb{\Phi}(\pmb{e})\pmb{h}=\pmb{h}^H\pmb{\Theta}\pmb{h},
\end{align}
where the codeword difference matrix can be found with the aid of \eqref{Eq28} and \eqref{Eq29}, which is given by
\begin{align}\label{Eq32}
\pmb{\Theta}=\begin{bmatrix}
	\pmb{e}^H\pmb{D}_1^H\pmb{D}_1\pmb{e} & \cdots & \pmb{e}^H\pmb{D}_1^H\pmb{D}_P\pmb{e}\\
	\vdots & \ddots & \vdots \\
	\pmb{e}^H\pmb{D}_P^H\pmb{D}_1\pmb{e} & \cdots & \pmb{e}^H\pmb{D}_P^H\pmb{D}_P\pmb{e}	
	\end{bmatrix}.
	\end{align}
The CPEP can be formulated as \cite{zhang2016compressed,zhang2018linear}
\begin{align}\label{Eq33}
P(\pmb{x}^c\rightarrow\pmb{x}^e|\pmb{h})&=P\left[||\pmb{y}-\pmb{\Phi}(\pmb{x}^e)\pmb{h}||_2^2\le||\pmb{y}-\pmb{\Phi}(\pmb{x}^c)\pmb{h}||_2^2\right]\nonumber\\
&=P[2\Re\left\{\pmb{y}^H\pmb{\Phi}(\pmb{x}^e)\pmb{h}\right\}-2\Re\left\{\pmb{y}^H\pmb{\Phi}(\pmb{x}^c)\pmb{h}\right\}\nonumber\\
&\quad\quad\geq||\pmb{\Phi}(\pmb{x}^e)\pmb{h}||_2^2-||\pmb{\Phi}(\pmb{x}^c)\pmb{h}||_2^2].
\end{align}
Upon exploiting \eqref{Eq28} and substituting $\pmb{\Phi}(\pmb{e})=\pmb{\Phi}(\pmb{x}^c)-\pmb{\Phi}(\pmb{x}^e)$ into \eqref{Eq33}, as well as following some rearrangement, \eqref{Eq33} can be further rewritten as
\begin{align}\label{EqA1}
P(\pmb{x}^c\rightarrow\pmb{x}^e|\pmb{h})=P\left[\Re\{\bar{\pmb{n}}^H\pmb{\Phi}(\pmb{e})\pmb{h}\}\geq\frac{\left \|\pmb{\Phi}(\pmb{e})\pmb{h}\right \|_2^2}{2}\right].
	\end{align}
It can be shown that $\Re\{\bar{\pmb{n}}^H\pmb{\Phi}(\pmb{e})\pmb{h}\}$ obeys Gaussian distribution with zero mean and a variance of $\frac{\left \|\pmb{\Phi}(\pmb{e})\pmb{h}\right \|_2^2}{2\gamma_s}$. Therefore, based on \eqref{Eq31}, the CPEP of \eqref{Eq33} can be expressed as
\begin{align}\label{EqA2}
	P(\pmb{x}^c\rightarrow\pmb{x}^e|\pmb{h})&=Q\left(\sqrt{\frac{\delta\gamma_s}{2}}\right)\leq\frac{1}{2}\exp\left(-\frac{\delta\gamma_s}{4}\right),
	\end{align}
where the inequality is obtained based on the well-known Chernoff bound $Q(x)\leq\frac{1}{2}\exp(-\frac{x^2}{2})$ \cite{tarokh1998space}. Explicitly, $\pmb{\Theta}$ is a positive semidefinite Hermitian matrix. Let $\text{rank}(\pmb{\Theta})=r$, where $1\leq r\leq P$. Then, when the eigenvalues are sorted in descending order and eigenvectors of $\pmb{\Theta}$ are denoted as $\left\{\lambda_1,\ldots,\lambda_r\right\}$ and $\left\{\pmb{\mu}_1,\ldots,\pmb{\mu}_r\right\}$, respectively, the CPEP can be further upper bounded by
\begin{align}\label{Eq34}
P(\pmb{x}^c\rightarrow\pmb{x}^e|\pmb{h})\leq\frac{1}{2}\exp\left(-\frac{\gamma_s\sum_{i=1}^r\lambda_i|\bar{h}_i|^2}{4}\right),
\end{align}
where $\bar{h}_i=\left\langle\pmb{h},\pmb{\mu}_i\right\rangle$, $\forall i\in[1,r]$, and we have $\bar{h}_i\sim\mathcal{CN}(v_i,1/P)$ with $v_i=\left\langle\mathbb{E}[\pmb{h}],\pmb{\mu}_i\right\rangle$. It can be readily shown that $|\bar{h}_i|$ are Rician distributed variables having the probability density function (PDF) of \cite{tarokh1998space}
\begin{align}\label{Eq35}
p(|\bar{h}_i|)=2P|\bar{h}_i|\exp(-P|\bar{h}_i|^2-P\zeta_i)I_0(2P|\bar{h}_i|\sqrt{\zeta_i}),
\end{align}
where $\zeta_i=|v_i|^2$ is the Rician factor. By averaging \eqref{Eq34} with respect to $|\bar{h}_i|$, the UPEP can be formulated as \cite{tarokh1998space}
\begin{align}\label{Eq36}
P(\pmb{x}^c\rightarrow\pmb{x}^e)\leq\frac{1}{2}\prod_{i=1}^r \frac{1}{1+\frac{\lambda_i\gamma_s}{4P}}\exp\left(-\frac{\frac{\zeta_i\lambda_i\gamma_s}{4P}}{1+\frac{\lambda_i\gamma_s}{4P}}\right).
\end{align}

Under the assumption of Rayleigh fading associated with $\zeta_i=0$, $|h_i|$ follows the Rayleigh distribution. Hence, \eqref{Eq36} can be expressed as
\begin{align}\label{Eq37}
P(\pmb{x}^c\rightarrow\pmb{x}^e)\leq\frac{1}{2\prod_{i=1}^r(1+\lambda_i\gamma_s/4P)}.
\end{align}
In the sufficiently high SNR regime $(\gamma_s\gg1)$, the UPEP can be formulated as
\begin{align}\label{Eq38}
	P(\pmb{x}^c\rightarrow\pmb{x}^e)\leq\frac{1}{2\prod_{i=1}^r\lambda_i}\left(\frac{\gamma_s}{4P}\right)^{-r}.
\end{align}

Finally, based on the union bounding technique, the BER of OTSM can be approximately written as
\begin{align}\label{Eq41}
P_e\approx\frac{1}{L_{b}2^{L_b}}\sum_{\pmb{x}^c}\sum_{\pmb{x}^e}P(\pmb{x}^c\rightarrow\pmb{x}^e)e(\pmb{x}^c,\pmb{x}^e),
\end{align}
where $e(\pmb{x}^c,\pmb{x}^e)$ denotes the corresponding number of different bits between $\pmb{x}^c$ and $\pmb{x}^e$. For high SNR values, the upper bound of the BER can be further simplified based on \eqref{Eq38} and \eqref{Eq41} as
\begin{align}\label{Eq42}
P_e\leq\frac{1}{2 L_{b}2^{L_b}}\sum_{\pmb{x}^c}\sum_{\pmb{x}^e}\frac{1}{\prod_{i=1}^r\lambda_i}\left(\frac{\gamma_s}{4P}\right)^{-r}e(\pmb{x}^c,\pmb{x}^e).
\end{align}
\vspace{-3em}
\subsection{Imperfect Channel Estimation Conditions}
By taking realistic channel estimation imperfections into account, the channel coefficient vector $\pmb{h}$ of \eqref{Eq28} can be rewritten as \cite{8933099}
\begin{align}\label{EqA3}
	\tilde{\pmb{h}}=\pmb{h}+\pmb{e}_h,
\end{align}
where $\pmb{e}_h$ represents the channel estimation error whose elements obeys $\mathcal{CN}(0,\sigma_h^2)$ with the variance $0\leq\sigma_h^2<1$. Under this condition, the input-output relationship of \eqref{Eq28} can be formulated as
\begin{align}\label{EqA4}
	\pmb{y}=\pmb{\Phi}(\pmb{x})\tilde{\pmb{h}}+\pmb{\Phi}(\pmb{x})(\pmb{h}-\tilde{\pmb{h}})+\bar{\pmb{n}}=\pmb{\Phi}(\pmb{x})\tilde{\pmb{h}}+\tilde{\pmb{n}},
\end{align}
where we have $\tilde{\pmb{n}}\triangleq\pmb{\Phi}(\pmb{x})(\pmb{h}-\tilde{\pmb{h}})+\bar{\pmb{n}}=-\pmb{\Phi}(\pmb{x})\pmb{e}_h+\bar{\pmb{n}}$. Therefore, similar to the derivation of \eqref{Eq33} and \eqref{EqA1}, the corresponding CPEP can be expressed as
\begin{align}\label{EqA5}
	P(\pmb{x}^c\rightarrow\pmb{x}^e|\tilde{\pmb{h}})&=P\left[||\pmb{y}-\pmb{\Phi}(\pmb{x}^e)\tilde{\pmb{h}}||_2^2\le||\pmb{y}-\pmb{\Phi}(\pmb{x}^c)\tilde{\pmb{h}}||_2^2\right]\nonumber\\
&=P[2\Re\left\{\pmb{y}^H\pmb{\Phi}(\pmb{x}^e)\tilde{\pmb{h}}\right\}-2\Re\left\{\pmb{y}^H\pmb{\Phi}(\pmb{x}^c)\tilde{\pmb{h}}\right\}\nonumber\\
&\quad\quad\geq||\pmb{\Phi}(\pmb{x}^e)\tilde{\pmb{h}}||_2^2-||\pmb{\Phi}(\pmb{x}^c)\tilde{\pmb{h}}||_2^2]\nonumber\\
&=P\left[\Re\{\tilde{\pmb{n}}^H\pmb{\Phi}(\pmb{e})\tilde{\pmb{h}}\}\geq\frac{\left \|\pmb{\Phi}(\pmb{e})\tilde{\pmb{h}}\right \|_2^2}{2}\right].
\end{align}
It can be readily observed that $\Re\{\tilde{\pmb{n}}^H\pmb{\Phi}(\pmb{e})\tilde{\pmb{h}}\}$ has a zero mean and variance of
\begin{align}\label{EqA6}
	Var\left[\Re\{\tilde{\pmb{n}}^H\pmb{\Phi}(\pmb{e})\tilde{\pmb{h}}\}\right]=\frac{\sigma_h^2\left \|\pmb{\Phi}(\pmb{x}^c)^H\pmb{\Phi}(\pmb{e})\tilde{\pmb{h}}\right \|_2^2+N_0\left \|\pmb{\Phi}(\pmb{e})\tilde{\pmb{h}}\right \|_2^2}{2}.
	\end{align}
Therefore, the CPEP of \eqref{EqA5} can be rewritten as
\begin{align}\label{EqA7}
	&P(\pmb{x}^c\rightarrow\pmb{x}^e|\tilde{\pmb{h}})\nonumber\\
	&=Q\left(\frac{\left \|\pmb{\Phi}(\pmb{e})\tilde{\pmb{h}}\right \|_2^2}{\sqrt{2\sigma_h^2\left \|\pmb{\Phi}(\pmb{x}^c)^H\pmb{\Phi}(\pmb{e})\tilde{\pmb{h}}\right \|_2^2+2N_0\left \|\pmb{\Phi}(\pmb{e})\tilde{\pmb{h}}\right \|_2^2}}\right).
\end{align}
Explicitly, the UPEP can be expressed as $P(\pmb{x}^c\rightarrow\pmb{x}^e)=\mathbb{E}_{\tilde{\pmb{h}}}\left[P(\pmb{x}^c\rightarrow\pmb{x}^e|\tilde{\pmb{h}})\right]$, and $\tilde{\pmb{h}}$ has the multivariate complex Gaussian PDF of
\begin{align}\label{EqA8}
	f(\tilde{\pmb{h}})=\frac{\pi^{-P}}{\det(\tilde{\pmb{\Psi}})}\exp\left(-\tilde{\pmb{h}}^H\tilde{\pmb{\Psi}}^{-1}\tilde{\pmb{h}}\right),
\end{align}
where $\tilde{\pmb{\Psi}}=\mathbb{E}\left\{\tilde{\pmb{h}}\tilde{\pmb{h}}^H\right\}=\pmb{\Psi}+\sigma_h^2\pmb{I}_P$ and $\pmb{\Psi}=\mathbb{E}\left\{{\pmb{h}}{\pmb{h}}^H\right\}$. However, given the complex structure of \eqref{EqA7}, calculating the UPEP directly is intractable. In the case of a constant envelope $K$-ary modulation constellation $\mathcal{A}$, it can be readily shown that $	\left \|\pmb{\Phi}(\pmb{x}^c)^H\pmb{\Phi}(\pmb{e})\tilde{\pmb{h}}\right \|_2^2\leq\left \|\pmb{\Phi}(\pmb{e})\tilde{\pmb{h}}\right \|_2^2$. Therefore, the upper-bound of the CPEP can be formulated as
\begin{align}\label{EqA10}
	P(\pmb{x}^c\rightarrow\pmb{x}^e|\tilde{\pmb{h}})=Q\left(\sqrt{\frac{\left \|\pmb{\Phi}(\pmb{e})\tilde{\pmb{h}}\right \|_2^2}{{2\sigma_h^2+2N_0}}}\right).
\end{align}
Based on \eqref{EqA8}, \eqref{EqA10} and the Chernoff bound $Q(x)\leq\frac{1}{2}\exp(-\frac{x^2}{2})$ \cite{tarokh1998space}, the UPEP upper-bound of the generalized OTSM system relying on realistic imperfect channel estimation can be derived as
\begin{align}\label{EqA11}
	P(\pmb{x}^c\rightarrow\pmb{x}^e)&\leq\frac{\pi^{-P}}{2\det(\tilde{\pmb{\Psi}})}\int_{\tilde{\pmb{h}}}\exp\left(-\tilde{\pmb{h}}^H\left[\tilde{\pmb{\Psi}}+\kappa\pmb{\Theta}\right]\tilde{\pmb{h}}\right)d\tilde{\pmb{h}}\nonumber\\
	&=\frac{1}{2\det(\pmb{I}_P+\kappa\tilde{\pmb{\Psi}}\pmb{\Theta})},
\end{align}
where $\kappa=1/(4\sigma^2_h+4N_0)$ and $\pmb{\Theta}=\pmb{\Phi(\pmb{e})}^H\pmb{\Phi(\pmb{e})}$. Finally, the BER upper bound under channel estimation imperfections can be calculated based on \eqref{Eq41}.
\section{Detection Algorithms for ZP-OTSM}\label{Section4}
In this section, we first discuss the optimum maximum \emph{a posteriori} (MAP) detector. Then, the family of linear detectors is introduced. Moreover, the conventional AMP detector is detailed. Furthermore, we propose our low-complexity VAMP-EM detector based on a factor graph. Finally, the complexity of different detectors is discussed.
\subsection{Maximum a posteriori Detection}\label{Section4-1}
In general, the optimal MAP detector of the OTSM system described in Section \ref{Section2} relies on maximizing the \emph{a posteriori} probability of $\pmb{x}$, given the received symbol vector $\pmb{y}$ as shown in \eqref{Eq18}, which is formulated as $\hat{\pmb{x}}^{\text{MAP}}=\underset{\pmb{x}\in\mathcal{A}^{MN}}{\arg\max} \{p(\pmb{x}|\pmb{y})\}$. Upon assuming that the mapping processes among different symbols are independent and equiprobable, the MAP detector can also be formulated as the ML detector equivalently of
\begin{align}\label{Eq44}
	\hat{\pmb{x}}^{\text{ML}}=\underset{\pmb{x}\in\mathcal{A}^{MN}}{\arg\min} \left\{||\pmb{y}-\pmb{H}\pmb{x}||_2^2\right\}.
\end{align}

Nevertheless, the complexity of the optimum ML detector is given by $\mathcal{O}(K^{MN})$, which is excessive for high values of $MN$. To mitigate this problem, it is crucial to design efficient detectors for large-scale OTSM systems.
\subsection{Linear Minimum Mean Square Error Detector}\label{Section4-2}
According to \eqref{Eq18}, the linear minimum mean square error (LMMSE) detector can be readily employed for recovering $\pmb{x}$, yielding,
\begin{align}\label{Eq44-1}
	\hat{\pmb{x}}^{\text{LMMSE}}=(\pmb{H}^H\pmb{H}+\frac{1}{\gamma_s}\pmb{I}_{MN})^{-1}\pmb{H}^H\pmb{y}.	
\end{align}
As shown in Fig. \ref{Figure3} \subref{Figure3-2}, the block circulant property cannot be satisfied by the ZP-OTSM system. Hence, the complexity of the LMMSE detector is still in the order of $\mathcal{O}(M^3 N^3)$, which is consistent with the observations in \cite{9285313}.

\subsection{Conventional AMP Detector}\label{Section4-3}
The AMP algorithm is initially proposed based on loopy BP amalgamated with Gaussian and Taylor series approximations, whose message passing process is conceived based on a scalar-valued factor graph \cite{donoho2009message}. Given the DS-domain input-output relationship in \eqref{Eq18}, the conventional AMP detector is formally stated in Algorithm \ref{alg1}. The AMP detector includes symbol denoising (SD) and LMMSE estimation (LE).

Given the modulation scheme, the \emph{a prior} information of $\pmb{x}$ can be written as $P(x(j)=a_k)=1/K$ for $j=0,\ldots,J-1$ and $k=1,\ldots,K$. The AMP algorithm quantities $\{\pmb{r},\pmb{x}\}$ obey
\begin{align}\label{Eq46}
	r(j)=x(j)+\omega(j),
\end{align}
where $\omega(j)\sim\mathcal{CN}[0,\upsilon_r(j)]$, and it can be observed that $\pmb{r}$ is viewed as the $\pmb{\upsilon_r}$-variance AWGN-contaminated version of the true signal $\pmb{x}$. In the SD part, lines 5-8 of Algorithm \ref{alg1} are leveraged to compute the $\emph{a posteriori}$ mean $x(j)$ and variance $\upsilon_x(j)$ of the elements in $\pmb{x}$. Moreover, the symbol denoising results in $\hat{\pmb{x}}$ can be regarded as an AWGN-contaminated version of the true symbol $\pmb{x}$, which is given by
 \begin{align}\label{Eq51}
	\hat{x}(j)=x(j)+\varpi(j),
\end{align}
where $\varpi(j)\sim\mathcal{CN}[0,\upsilon_x(j)]$. In the LE part, lines 10-15 of Algorithm \ref{alg1} are directly given by the AMP algorithm to calculate $r(j)$. 
The term $\pmb{\upsilon}_r^{t+1}\cdot\pmb{H}^H \pmb{s}^{t+1}$ in line 15 of Algorithm \ref{alg1} is the Onsager term of \cite{donoho2009message} that asymptotically decouples the denoising part and the LMMSE estimation part through iterations, i.e., eliminates the correlation between these two parts.
\begin{algorithm}[htbp]
\footnotesize
\caption{AMP Detector}
\label{alg1}
\begin{algorithmic}[1]
    \Require
      $\pmb{y}$, $\pmb{H}$ and $\gamma_n$.
      \State \textbf{Preparation}: Set a maximum iterations number $T_{\text{AMP}}$ and an error tolerance parameter $\epsilon$. 
    \State \textbf{Initialize} $\pmb{s}^{(1)}=\pmb{0}$, $\pmb{\upsilon}_r^{(1)}=\pmb{1}$ and $\pmb{r}^{(1)}=\pmb{0}$.
    \For{$t=1$ to $T_{\text{AMP}}$}
    \State $// \text{Symbol denoising}$
    \State $\forall j,k:$ $\xi_{j,k}^t=\exp\left[-|a_k-r^t(j)|^2/\upsilon_{r}(j)^t\right]$
    \State $\forall j,k:$ $\beta_{j,k}=\xi_{j,k}^t/\sum_{k=1}^{K}\xi_{j,k}^t$
    \State $\forall j:$ $\hat{x}^t(j)=\sum_{k=1}^{K}a_k\beta_{j,k}$
    \State $\forall j:$ $\upsilon_x^t(j)=\sum_{k=1}^{K}\beta_{j,k}|a_k-\hat{x}^t(j)|^2$
    \State $// \text{LMMSE estimation}$
    \State $\pmb{\upsilon}_p=|\pmb{H}|^2\pmb{\upsilon}_x^t$
    \State $\pmb{p}=\pmb{H}\hat{\pmb{x}}^t-\pmb{\upsilon}_p\cdot\pmb{s}^{t}$
    \State $\pmb{\upsilon}_{s}=\pmb{1}\cdot/\left(\pmb{\upsilon}_{p}+\gamma_n^{-1} \pmb{1}\right)$
    \State $\pmb{s}^{t+1}=\pmb{\upsilon}_s\cdot(\pmb{y}-\pmb{p})$
    \State $\pmb{\upsilon}_r^{t+1}=\pmb{1}\cdot/|\pmb{H}^H|^2\pmb{\upsilon}_s$
    \State $\pmb{r}^{t+1}=\hat{\pmb{x}}^t+\pmb{\upsilon}_r^{t+1}\cdot\pmb{H}^H \pmb{s}^{t+1}$
    \State \textbf{if} $\|\hat{\pmb{x}}^{t+1}-\hat{\pmb{x}}^{t}\|_2^2<\epsilon\|\hat{\pmb{x}}^{t}\|_2^2$
    \State \textbf{break}
\EndFor
\State \textbf{return} $\hat{\pmb{x}}$.
\end{algorithmic}
\end{algorithm}

It should be noted that the AMP detector has two limitations. First, AMP is designed for accommodating zero-mean i.i.d Gaussian sensing matrices \cite{donoho2009message}. Bearing in mind the specific structure of the DS-domain channel matrix $\pmb{H}$ shown in Fig. \ref{Figure3} \subref{Figure3-2}, the performance of the AMP detector may degrade significantly or even diverge. Moreover, the AMP detector proceeds iteratively with the known noise variance $1/\gamma_n$, which is however rarely known in practice. To alleviate the above issues, we propose the VAMP-EM detector in the next subsection.

\subsection{VAMP-EM Detector}\label{Section4-4}
The VAMP algorithm is derived from a non-loopy factor graph having vector-valued nodes, which is different from AMP \cite{rangan2019vector}. Hence, VAMP achieves better MSE performance than AMP in the face of ill-conditioned sensing matrices. By intrinsically amalgamating the VAMP and the EM algorithms, we propose the VAMP-EM detector summarized in Algorithm \ref{alg2}, which includes two inner iterations and one outer iterations. Specifically, we consider three aspects, which are detailed below.
\begin{figure}[htbp]
\centering
\includegraphics[width=\linewidth]{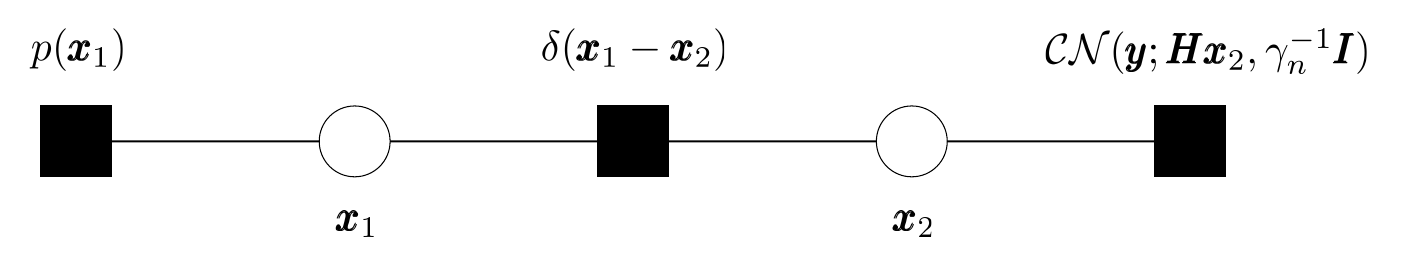}
\caption{Illustration of the VAMP-EM factor graph}
\label{Figure4}
\end{figure}
\subsubsection{VAMP Algorithm}
We first factorise the joint likelihood function of $\pmb{y}$ and $\pmb{x}$ formulated as $p(\pmb{y},\pmb{x})=p(\pmb{x})\mathcal{CN}(\pmb{y};\pmb{H}\pmb{x}, \pmb{I}/\gamma_n)$. As shown in Fig. \ref{Figure4}, based on the derivation of the VAMP illustrated in \cite{rangan2019vector}, the transmitted symbol vector $\pmb{x}$ is split into two variables $\pmb{x}_1=\pmb{x}_2$. Hence, the equivalent factorization of $p(\pmb{y},\pmb{x})$ can be further expressed as $p(\pmb{y},\pmb{x}_1,\pmb{x}_2)=p(\pmb{x}_1)\delta(\pmb{x}_1-\pmb{x}_2)\mathcal{CN}(\pmb{y};\pmb{H}\pmb{x}_2, \pmb{I}/\gamma_n)$. In the $t$th outer loop iteration, the message passing process commences with $\mu_{\delta\rightarrow\pmb{x}_1}(\pmb{x}_1)=\mathcal{CN}\left[\pmb{x}_1;\pmb{r}_1^t,(\gamma_1^t\pmb{I})^{-1}\right]$. Then the approximate belief of $\pmb{x}_1$ can be calculated as $b_{\text{app}}(\pmb{x}_1)=\mathcal{CN}\left[\pmb{x}_1;\hat{\pmb{x}}_1^t,(\eta_1^t)^{-1}\pmb{I}\right]$.

In the SD part, based on the statistics of the modulation scheme, the probability density function of $\pmb{x}$ can be formulated as $p(\pmb{x})=\prod_{j=1}^{MN}p(x_j)=\prod_{j=1}^{MN}\frac{1}{K}\sum_{k=1}^K \delta(x_j-a_k)$. Similar to AMP, we obtain the scalar-valued $\emph{a posteriori}$ mean $\hat{x}_1(j)$ and variance $\upsilon_{x_{1}}(j)$,  yielding lines 6-9 of Algorithm \ref{alg2}. Thus the average variance can be written as $1/\eta_1^\tau=\frac{1}{MN}\sum_{j=1}^{MN}\upsilon_{x_{1}}(j)$, yielding line 10 of Algorithm \ref{alg2}. Hence we obtain $b_{\text{app}}(\pmb{x}_1)$. In the classic VAMP algorithm, under the assumption that $\pmb{H}$ is right-rotationally invariant, both $\hat{\pmb{x}}_1$ and $\hat{\pmb{x}}_2$ can be obtained by LMMSE estimation. Then the outputs of the SD and LE parts obey \cite{rangan2019vector}
\begin{align}\label{Eq56}
	\pmb{r}_1^t=\pmb{x}+\mathcal{CN}(0,\pmb{I}/\gamma_1^t),\nonumber\\
	\pmb{x}=\pmb{r}_2^t+\mathcal{CN}(0,\pmb{I}/\gamma_2^t).
	\end{align}
Hence, $\pmb{r}_1^t$ may be viewed as a $1/\gamma_1^t$-variance AWGN-contaminated version of the true signal $\pmb{x}$, and the true symbol vector $\pmb{x}$ can be regarded as a $1/\gamma_2^t$-variance AWGN-contaminated version of $\pmb{r}_2^t$. Based on the classic message passing rules of VAMP \cite{rangan2019vector}, the message $\mu_{\pmb{x}_1\rightarrow\delta}(\pmb{x}_1)=\mathcal{CN}\left[\pmb{x}_1;\pmb{r}_2^t,(\gamma_2^t)^{-1}\pmb{I}\right]$ can be formulated as
\begin{align}\label{Eq57}
	\mu_{\pmb{x}_1\rightarrow\delta}(\pmb{x}_1)&=\mathcal{CN}\left[\pmb{x}_1;\hat{\pmb{x}}_1^t,(\eta_1^t)^{-1}\pmb{I}\right]/\mathcal{CN}(\pmb{x}_1;\pmb{r}_1^t,(\gamma_1^t)^{-1}\pmb{I})\nonumber\\
	&\propto\mathcal{CN}\left[\pmb{x}_1;(\hat{\pmb{x}}_1^t\eta_1^t-\pmb{r}_1^t\gamma_1^t)/(\eta_1^t-\gamma_1^t),(\eta_1^t-\gamma_1^t)^{-1}\right].
\end{align}
Therefore, we can obtain $\pmb{r}_2^t=(\eta_1^t\hat{\pmb{x}}_1^t-\gamma_1^t\pmb{r}_1^t)/\gamma_2^t$ and $\gamma_2^t=\eta_1^t-\gamma_1^t$, yielding lines 14-15 of Algorithm \ref{alg2}. It can be readily shown in Fig. \ref{Figure4} that the message $\mu_{\pmb{x}_1\rightarrow\delta}(\pmb{x}_1)=\mathcal{CN}\left[\pmb{x}_1;\pmb{r}_2^t,(\gamma_2^t)^{-1}\pmb{I}\right]$ remains constant after passing through the node $\delta$, hence we have $\mu_{\delta\rightarrow\pmb{x}_2}(\pmb{x}_2)=\mu_{\pmb{x}_1\rightarrow\delta}(\pmb{x}_1)=\mathcal{CN}\left[\pmb{x}_1;\pmb{r}_2^t,(\gamma_2^t)^{-1}\pmb{I}\right]$.

In the LE part, according to the factor graph message passing rules \cite{rangan2019vector}, the approximate belief of the node $\pmb{x}_2$ in Fig. \ref{Figure4} can be expressed as $b_\text{app}(\pmb{x}_2)=\mathcal{CN}\left[\pmb{x}_2,\hat{\pmb{x}}_2^t,(\gamma_2^t)^{-1}\pmb{I}\right]$, where $\hat{\pmb{x}}_2^t=\mathbb{E}[\pmb{x}_2|b_\text{sp}(\pmb{x}_2)]$ and $\eta_2^t=\left<\text{diag}(\text{Cov}\left[\pmb{x}_2|b_{\text{sp}}(\pmb{x}_2)\right])\right>^{-1}$, the sum-product (SP) belief concerning $\pmb{x}_2$ can be formulated as $b_{\text{sp}}(\pmb{x}_2)\propto\mathcal{CN}\left[\pmb{x}_2;\pmb{r}_2^t,(\eta_2^t)^{-1}\pmb{I}\right]\mathcal{CN}\left[\pmb{y};\pmb{H}\pmb{x}_2,(\gamma_n)^{-1}\pmb{I}\right]$. Consequently, the conditional-mean and covariance of $b_{\text{app}}(\pmb{x}_2)$ can be formulated as \cite{rangan2019vector}
\begin{align}\label{Eq59-1}
	\hat{\pmb{x}}_2^t=\left(\gamma_n^\tau\pmb{H}^H\pmb{H}+\gamma_2^\tau\pmb{I}\right)^{-1}\left(\gamma_n^\tau\pmb{H}^H\pmb{y}+\gamma_2^\tau\pmb{r}_2^\tau\right),
	\end{align}
\begin{align}\label{Eq60-1}
	(\eta_2^t)^{-1}\pmb{I}=\left(\gamma_n^\tau\pmb{H}^H\pmb{H}+\gamma_2^\tau\pmb{I}\right)^{-1}.
	\end{align}
Hence, upon exploiting
\begin{align}\label{Eq59}
	\pmb{g}_2(\pmb{r}_2^\tau,\gamma_2^\tau)=\left(\gamma_n^\tau\pmb{H}^H\pmb{H}+\gamma_2^\tau\pmb{I}\right)^{-1}\left(\gamma_n^\tau\pmb{H}^H\pmb{y}+\gamma_2^\tau\pmb{r}_2^\tau\right),
	\end{align}
\begin{align}\label{Eq60}
	\left\langle\pmb{g}_2'(\pmb{r}_2^\tau,\gamma_2^\tau)\right \rangle=\gamma_2^\tau\text{tr}\left\{\left(\gamma_n^\tau\pmb{H}^H\pmb{H}+\gamma_2^\tau\pmb{I}\right)^{-1}\right\}/MN,
	\end{align}
lines 18-20 of Algorithm \ref{alg2} have been interpreted.

In order to avoid the matrix inversion process in \eqref{Eq59} and \eqref{Eq60}, we apply the singular value decomposition (SVD) to the DS-domain DT channel matrix, yielding $\pmb{H}=\pmb{U}\pmb{S}\pmb{V}^H$, where we have $\pmb{S}=\text{diag}[s_1,\ldots,s_{MN}]$ and $R=\text{rank}(\pmb{H})$. Hence, $\pmb{g}_2(\pmb{r}_2^\tau,\gamma_2^\tau)$ and $\left\langle\pmb{g}_2'(\pmb{r}_2^\tau,\gamma_2^\tau)\right \rangle$ may be written as
\begin{align}\label{Eq61}
	\pmb{g}_2(\pmb{r}_2^\tau,\gamma_2^\tau)=\pmb{V}\pmb{\Xi}^\tau(\bar{\pmb{y}}+\gamma_2^\tau\pmb{V}^H\pmb{r}_2^\tau),
	\end{align}
\begin{align}\label{Eq62}
	\left\langle\pmb{g}_2'(\pmb{r}_2^\tau,\gamma_2^\tau)\right \rangle=\frac{1}{MN}\sum_{j=1}^{MN}\frac{\gamma_2^\tau}{\gamma_n|s_j|^2+\gamma_2^\tau},
	\end{align}
where $\bar{\pmb{y}}=\gamma_n\pmb{S}^H\pmb{U}^H\pmb{y}$ and $\pmb{\Xi}^\tau\in\mathbb{R}^{MN\times MN}$ denotes a diagonal matrix associated with $\Xi^\tau(j,j)=(\gamma_n|s_j|^2+\gamma_2^\tau)^{-1}$. Then we obtain the \emph{a posteriori} mean symbol vector $\hat{\pmb{x}}_2^t$ and the variance $1/\eta_2^t$. It maybe readily shown that the message $\mu_{\pmb{x}_2\rightarrow\delta}(\pmb{x}_2)$ can be formulated as \cite{rangan2019vector}
\begin{align}
	\mu_{\pmb{x}_2\rightarrow\delta}(\pmb{x}_2)&=\mathcal{CN}\left[\pmb{x}_2;\pmb{r}_1^{t+1},(\gamma_1^{t+1})^{-1}\pmb{I}\right]\nonumber\\
&=\mathcal{CN}\left[\pmb{x}_2;\hat{\pmb{x}}_2^t,(\eta_2^t)^{-1}\pmb{I}\right]/\mathcal{CN}\left[\pmb{x}_2;\pmb{r}_2^t,(\gamma_2^t)^{-1}\right].
\end{align}
Consequently, we can obtain $\pmb{r}_1^{t+1}=(\eta_2^t\hat{\pmb{x}}_2^t-\gamma_2^t\pmb{r}_2^t)/(\eta_2^t-\gamma_2^t)$ and $\gamma_1^{t+1}=\eta_2^t-\gamma_2^t$. It should be noted that the DS-domain channel matrix $\pmb{H}$ cannot satisfy the right-rotationally invariant property. Hence, for the ill-conditioned $\pmb{H}$, the performance of the VAMP-EM detector may be poor. To circumvent this impediment, a damping strategy is applied to the VAMP-EM detection algorithm \cite{9507331,8712432}. This justifies lines 25-26 of Algorithm \ref{alg2}, where the damping factor obeys $\theta\in(0,1]$. In general, the optimal $\theta$ can be obtained for a given SNR by searching within the interval $\theta\in(0:0.1:1]$ based on the minimum BER criteria. Since the message through the node $\delta$ remains unchanged, we have $\mu_{\delta\rightarrow\pmb{x}_1}(\pmb{x}_1)=\mathcal{CN}\left[\pmb{x}_1;\pmb{r}_1^{t+1},(\gamma_1^{t+1})^{-1}\right]$. Then the above message passing iterations are repeated with $t\leftarrow t+1$.

Compared to AMP, we noticed that both the AMP and VAMP include the SD and LE parts, hence the denoising steps in lines 5-8 of Algorithm \ref{alg1} are equivalent to lines 6-9 of Algorithm \ref{alg2}. But the outputs of the SD part within the AMP detector are the \emph{a posteriori} mean $\pmb{x}$ and variance $\pmb{\upsilon_x}$ without Onsager correction, which leads to correlations between $\pmb{r}$ as well as $\pmb{x}$, and to a loss of performance.
\begin{algorithm}[htbp]
\footnotesize
\caption{VAMP-EM Detector}
\label{alg2}
\begin{algorithmic}[1]
    \Require
      $\pmb{y}$, $\pmb{H}=\pmb{U}\pmb{S}\pmb{V}^H$ with $\pmb{S}=\text{diag}[s_1,\ldots,s_{MN}]$ and $R=\text{rank}(\pmb{H})\leq MN$.
      \State \textbf{Preparation}: Set a maximum iterations number $T$, a damping factor $\theta$ and an error tolerance parameter $\epsilon$. 
    \State \textbf{Initialize} $\gamma_1^{(1)}=0$, $1/\gamma_n^{(1)}=||\pmb{y}||_2^2/MN$ and $\pmb{r}_1^{(1)}=\pmb{0}$.
    \For{$t=1$ to $T$}
    \State $// \text{Symbol denoising}$
    \For{$\tau=1$ to $T_1$}    
    \State $\forall j,k:$ $\xi_{j,k}^\tau=\exp\left[-\gamma_1^\tau|a_k-r_1^\tau(j)|^2\right]$
    \State $\forall j,k:$ $\beta_{j,k}^\tau=\xi_{j,k}^\tau/\sum_{k=1}^{K}\xi_{j,k}^\tau$
    \State $\forall j:$ $\hat{x}_1^\tau(j)=\sum_{k=1}^{K}a_k\beta_{j,k}^\tau$
    \State $\forall j:$ $\upsilon_{x_{1}}(j)=\sum_{k=1}^{K}\beta_{j,k}^\tau|a_k-\hat{x}_1^\tau(j)|^2$
    \State $1/\eta_1^{\tau}=\frac{1}{MN}\sum_{j=1}^{MN}\upsilon_{x_{1}}(j)$
    \State $1/\gamma_1^{\tau+1}=\frac{1}{MN}||\hat{\pmb{x}}_1-\pmb{r}_1^\tau||_2^2+1/\eta_1^{\tau}$
    \EndFor
    \State $\hat{\pmb{x}}_1^t=\hat{\pmb{x}}_1^{T_1}$, $1/\eta_1^t=1/\eta_1^{T_1}$, $1/\gamma_1^t=1/\gamma_1^{T_1+1}$    
    \State $\gamma_2^t=\eta_1^t-\gamma_1^t$
    \State $\pmb{r}_2^t=(\eta_1^t\hat{\pmb{x}}_1^t-\gamma_1^t\pmb{r}_1^t)/\gamma_2^t$
    \State $// \text{LMMSE estimation}$
    \For{$\tau=1$ to $T_2$}
    \State $\hat{\pmb{x}}_2^\tau=\pmb{g}_2(\pmb{r}_2^\tau,\gamma_2^\tau)$
    \State $\alpha^\tau_2=\left\langle\pmb{g}_2'(\pmb{r}_2^\tau,\gamma_2^\tau)\right \rangle$
    \State $1/\eta_2^\tau=\alpha^\tau_2/\gamma_2^\tau$
     \State $1/\gamma_2^{\tau+1}=\frac{1}{MN}||\hat{\pmb{x}}_2^\tau-\pmb{r}_2^\tau||_2^2+1/\eta_2^\tau$
    \State $1/\gamma_n^{\tau+1}=\frac{1}{MN}\left[||\pmb{y}-\pmb{H}\hat{\pmb{x}}_2^\tau||_2^2+\sum_{k=1}^{R}\frac{|s_k|^2}{\gamma_n^\tau |s_k|^2+\gamma_2^{\tau+1}}\right]$
    \EndFor
    \State $\hat{\pmb{x}}_2^t=\hat{\pmb{x}}_2^{T_2}$, $1/\eta_2^t=1/\eta_2^{T_2+1}$, $1/\gamma_2^t=1/\gamma_2^{T_2+1}$
    \State $\gamma_1^{t+1}=(1-\theta)\gamma_1^t+\theta(\eta_2^t-\gamma_2^t)$
    \State $\pmb{r}_1^{t+1}=(1-\theta)\pmb{r}_1^{t}+\theta(\eta_2^t\hat{\pmb{x}}_2^t-\gamma_2^t\pmb{r}_2^t)/(\eta_2^t-\gamma_2^t)$    
    \State \textbf{if} $\|\hat{\pmb{x}}^{t+1}_1-\hat{\pmb{x}}_1^{t}\|_2^2<\epsilon\|\hat{\pmb{x}}^{t}_1\|_2^2$
    \State \textbf{break}
\EndFor
\State \textbf{return} $\hat{\pmb{x}}_1$.
\end{algorithmic}
\end{algorithm}
\subsubsection{Learning $\gamma_n$}
Now we consider the problem of estimating the noise variance $1/\gamma_n$. Given the DS-domain input-output relationship shown in \eqref{Eq18}, the likelihood function of $\pmb{y}$ can be written as
\begin{align}\label{Eq67}
p(\pmb{y}|\pmb{x};\gamma_n)=\left(\frac{\gamma_n}{\pi}\right)^{MN}\exp\left[-\gamma_n||\pmb{y}-\pmb{H}\pmb{x}||_2^2\right].	\end{align}
Consequently, the joint probability density function of $\pmb{x}$ and $\pmb{y}$ given $\gamma_n$ can be formulated as $p(\pmb{x},\pmb{y};\gamma_n)=p(\pmb{x})p(\pmb{y}|\pmb{x};\gamma_n)$. Then the EM algorithm is employed to obtain the ML estimate of the unknown hyper-parameters by iteratively minimizing the negative likelihood upper bound and tightening it \cite{neal1998view}. The process of the EM algorithm can be formulated as
\begin{align}\label{Eq69}
V\left(\gamma_n; \hat{\gamma}^{t}_n\right) \triangleq-\mathbb{E}\left[\ln p(\boldsymbol{x}, \boldsymbol{y} ; \gamma_n) \mid \boldsymbol{y} ; \hat{\gamma}^{t}_n\right],
	\end{align}
\begin{align}\label{Eq70}
	\hat{\gamma}^{t+1}_n=\arg \min _{\gamma_n} V\left(\gamma_n; \hat{\gamma}^{t}_n\right).
	\end{align}
The generalized EM framework based on Gibbs sampling is detailed in \textbf{Appendix \ref{Appendix}}. To compute \eqref{Eq69} and \eqref{Eq70}, we consider the Gibbs energy function of \cite{vila2013expectation}
\begin{align}\label{Eq76}
J(b,q;\gamma_n)=D_\text{KL}(b,\gamma_n)+H(q),
\end{align}
which is equivalent to \eqref{Eq74}, when $b=q$. Hence the minimization of the negative likelihood upper bound can be rewritten as
\begin{align}\label{Eq77}
&\hat{\gamma}^{t+1}_n=\arg\min_{\gamma_n}\min_b\max_{q}J\left(b,q, \gamma_n\right);\nonumber\\
&\text{s.t.}\ \mathbb{E}\left[\boldsymbol{x}|b\right]=\mathbb{E}[\boldsymbol{x}|q],\quad\operatorname{tr}\left\{\operatorname{Cov}\left[\boldsymbol{x}|b\right]\right\}=\operatorname{tr}\{\operatorname{Cov}[\boldsymbol{x}|q]\}.
\end{align}
According to \textbf{Appendix \ref{Appendix}}, $\gamma_n$ can be estimated by solving the following optimization:
\begin{align}\label{Eq78}
	\hat{\gamma}_{n}^{t+1}=\arg \max _{\gamma_n} \mathbb{E}\left[\ln p\left(\boldsymbol{y}|\boldsymbol{x},\gamma_n\right)|\boldsymbol{r}_{2}^{t},\gamma_{2}^{t},\hat{\gamma}_{n}^{t}\right].
\end{align}
Specifically, based on the Gaussian likelihood function \eqref{Eq67}, we obtain $\hat{\gamma}_{n}^{t+1}$ as 
\begin{align}\label{Eq79}
	1/\hat{\gamma}_{n}^{t+1}&=\frac{1}{MN}\mathbb{E}\left[\|\boldsymbol{y}-\pmb{H}\boldsymbol{x}\|_2^{2}|\boldsymbol{r}_{2}^{t}, \gamma_{2}^{t},\gamma_{n}^{t}\right]\nonumber\\
	&=\frac{1}{MN}\left[||\pmb{y}-\pmb{H}\pmb{x}_2^t||_2^2+\sum_{k=1}^{R}\frac{|s_k|^2}{\gamma_n^t |s_k|^2+\gamma_2^t}\right],
\end{align}
thus interpreting line 22 of Algorithm \ref{alg2}.

\subsubsection{Adaptive Auto-tuning}
Under the condition that the input-output relationship \eqref{Eq18} relying on imperfectly estimated noise variance $1/\gamma_n$, the statistical model of VAMP quantities shown in \eqref{Eq56} cannot be satisfied. Specifically, in this case, $\pmb{r}_1^t$ is obtained as an AWGN-contaminated version of the true signal $\pmb{x}$ associated with another AWGN precision value based on the state evolution function \cite{rangan2019vector}. Therefore, the values of $\pmb{r}_i^t$ are characterized by $\gamma_i^t$ imprecisely hence resulting in the imperfect estimation of $\gamma_n$. In order to mitigate this problem, inspired by the adaptive VAMP algorithm \cite{DBLP:journals/corr/FletcherSSR17}, the re-estimation process of the noise precision $\left\{\gamma_i^t\right\}$ is applied to the VAMP-EM detector. Based on \eqref{Eq56}, the ML estimation of $\gamma_1^t$ can be formulated as $\gamma_1^t=\argmax_{\gamma_1}p(\pmb{r}_1^t;\gamma_1)$. By employing the EM algorithm, the above problem can be solved indexed by $\tau$, yielding
\begin{align}\label{Eq81}
	\gamma_1^{\tau+1}&=\argmax_{\gamma_1}\mathbb{E}\left[\ln p(\pmb{x},\pmb{r}_1^{\tau};\gamma_1)|\pmb{r}_1^{\tau};\gamma_1^{\tau}\right]\nonumber\\
	&=\argmax_{\gamma_1}\mathbb{E}\left[\ln p(\pmb{r}_1^{\tau}|\pmb{x};\gamma_1)|\pmb{r}_1^{\tau};\gamma_1^{\tau}\right].
	\end{align}
Therefore, the iterations of $\gamma_1^{\tau}$ can be further expressed as
\begin{align}\label{Eq82}
	\gamma_1^{\tau+1}&=\argmax_{\gamma_1}\left\{MN\ln\gamma_1-\gamma_1\mathbb{E}\left[||\pmb{x}-\pmb{r}_1^t||_2^2|\pmb{r}_1^t;\gamma_1^{\tau}\right]\right\}\nonumber\\
	&=MN\left\{\mathbb{E}\left[||\pmb{x}-\pmb{r}_1^t||_2^2|\pmb{r}_1^t;\gamma_1^{\tau}\right]\right\}^{-1}\nonumber\\
	&=\left\{\frac{1}{MN}\sum_{j=1}^{MN}\mathbb{E}[|x(j)-r_{1}(j)|^2|\pmb{r}_1^{\tau};\gamma_1^{\tau}]\right\}^{-1}\nonumber\\
	&=\left\{\frac{1}{MN}||\pmb{x}-\pmb{r}_1^{\tau}||_2^2+\frac{1}{\eta_1^{\tau}}\right\}^{-1}.
\end{align}
Finally, the re-estimation process of $\gamma_1^t$ is detailed above, and the re-estimation result of $\gamma_2^t$ can be attained similarly. Hence, by employing the EM algorithm, we carry out the first $T_1$-iterations within the inner loop, while the second inner loop contains $T_2$ iterations, yielding lines 5-12 and 17-23 of Algorithm \ref{alg2}.
\vspace{-1.5em}
\subsection{Complexity Analysis}\label{Section4-5}
The complexity of the single-tap detector is on the order of $\mathcal{O}(MN\log_2 M)$ \cite{thaj2021orthogonal}. Moreover, the complexity order of the GS detector is given by $\mathcal{O}(M^2 NL+M^2 NT_\text{GS})$ \cite{thaj2021orthogonal}. Furthermore, the complexity of the unitary AMP (UAMP) detector can be formulated as $\mathcal{O}(M^2NT_\text{UAMP}+MNQT_\text{UAMP})$ \cite{yuan2021iterative}.

Based on the analysis in Section \ref{Section4-3}, the SD part of the conventional AMP detector has a complexity order given by $\mathcal{O}(MNQT_{\text{AMP}})$, while the LE part's complexity is on the order of $\mathcal{O}(MN^2LT_{\text{AMP}})$, which takes advantage of the sparse structure of $\pmb{H}$. Hence, the overall complexity of the AMP detector can be expressed as $\mathcal{O}(MNQT_{\text{AMP}}+MN^2LT_{\text{AMP}})$.

Finally, we consider the VAMP-EM detector of Section \ref{Section4-4}. Similarly, the VAMP-EM detector complexity of the SD part in every iteration is given by $\mathcal{O}(MNQT_1)$, while that of the LE part in every iteration is on the order of $\mathcal{O}(M^2 N^2T_2)$. Therefore, the overall complexity of the VAMP-EM detector can be expressed as $\mathcal{O}[(MNQT_1+M^2 N^2T_2)T]$.
\vspace{-1.5em}
\section{Turbo Receiver of The Coded OTSM System}\label{Section5}
In this section, we first discuss the LDPC-coded OTSM system, then detail the structure of both the AMP and VAMP-EM turbo receivers.

We consider the LDPC-coded OTSM system shown in Fig. \ref{Figure1}. The turbo receiver includes a soft-decision detector, deinterleaver, and soft LDPC decoder. In the receiver, the soft-information of detected symbols is iteratively exchanged between these three blocks in the form of extrinsic log-likelihood ratios (LLRs). The extrinsic LLRs output by the symbol-to-bit converter (SBC) can be expressed as \cite{tuchler2002turbo}
\begin{align}\label{Eq85}
	L_e^E(c_{j}(n))&=\ln\frac{P(c_{j}(n)=0|\pmb{y})}{P(c_{j}(n)=1|\pmb{y})}-L_a^D(c_{j}(n))\nonumber\\
		&=\ln\frac{\sum\limits_{\forall{a_k: s_k(n)=0}}\varrho_{j,k}\prod\limits_{\forall n':n'\neq n}P(c_{j}(n')=s_k(n'))}{\sum\limits_{\forall{a_k: s_k(n)=1}}\varrho_{j,k}\prod\limits_{\forall n':n'\neq n}P(c_{j}(n')=s_k(n'))},
\end{align}
where $\varrho_{j,k}=\exp\left(-\frac{|\chi(j)-a_k|^2}{\sigma(j)}\right)$ and $L_a^D(c_{j}(n))=P(c_{j}(n)=0)/P(c_{j}(n)=1)$ represents the \emph{a priori} LLRs obtained from the interleaver, for $j=0,\ldots,J-1$ and $n=1,\ldots,\log_2 K$. Furthermore, $\chi(j)$ and $\sigma(j)$ represent the extrinsic mean and variance of $x(j)$, respectively. In the AMP turbo receiver, the output of Algorithm \ref{alg1} can be modified as \cite{yuan2021iterative}
\begin{align}\label{Eq86}	
\chi(j)=r(j),\quad\sigma(j)=\upsilon_r(j).
\end{align}
Let us now consider Algorithm \ref{alg2}. As shown in Fig. \ref{Figure4}, based on the derivation of the VAMP algorithm \cite{rangan2019vector} and lines 25-26 of Algorithm \ref{alg2}, it can be found that $\pmb{r}_1$ and $\gamma_1$ are attained based on the message passed from the variable node $\pmb{x}_2$ to the factor node $\delta(\pmb{x}_1-\pmb{x}_2)$, which is independent of the \emph{a priori} message gleaned from the variable node $\pmb{x}_1$. Hence, $\pmb{r}_1$ and $\gamma_1$ can be regarded as the the output of Algorithm \ref{alg2}, yielding
\begin{align}\label{Eq87}	
\chi(j)=r_1(j),\quad\sigma(j)=\gamma_1.
\end{align}
Moreover, for the symbol denosing parts of turbo receivers, the \emph{a priori} symbol input probability matrix $\pmb{P}$ can be provided based on $L_a^D(c_{j}(n))$ from the bit-to-symbol converter (BSC). Therefore, line 5 of Algorithm \ref{alg1} is modified as
\begin{align}\label{Eq88}	
\xi_{j,k}^t=P(x(j)=a_k)\exp\left[-|a_k-r^t(j)|^2/\upsilon_{r}(j)^t\right],
\end{align}
and line 6 of Algorithm \ref{alg2} can be amended as
\begin{align}\label{Eq89}	
\xi_{j,k}^\tau=P(x(j)=a_k)\exp\left[-\gamma_1^\tau|a_k-r_1^\tau(j)|^2\right].
\end{align}
where $P(x(j)=a_k)$ represent the elements of $\pmb{P}$. Furthermore, the probability $P(x(j)=a_k)$ can be formulated as \cite{tuchler2002turbo}
\begin{align}\label{Eq90}	
P(x(j)=a_k)&=\prod_{n=1}^{\log_2 K}P(c_{j}(n)=s_k(n))\nonumber\\
&=\prod_{n=1}^{\log_2 K}1/2\left[1+\tilde{s}_k(n)\tanh(L_a^D(c_{j}(n)/2))\right],
\end{align}
where
\begin{align}\label{Eq91}	
\tilde{s}_k(n)=
\begin{cases}
	+1, & s_k(n)=0\\
	-1, & s_k(n)=1.
	\end{cases}
\end{align}
\section{Numerical Results}\label{Section6}
In this section, simulation results are provided for characterizing
the overall attainable performance of the proposed detectors in both uncoded and coded OTSM systems.
\begin{table}[htbp]
\vspace{-1.5em}
\footnotesize
\centering
\caption{Simulation parameters for BER performance analysis}
\label{table2}
\begin{tabular}{l|l}
\hline
\textbf{Parameters} & \textbf{Values} \\
\hline
\hline
Maximum delay-domain grid index, $M$ & 2 \\  
\hline
No. of OTSM symbol, $N$ & 2, 4 \\
\hline
Carrier frequency, $f_c$ & 4 GHz\\
\hline
Subcarrier spacing, $\Delta f$ & 3.75 kHz\\
\hline
No. of paths, $P$ & 4\\
\hline
Maximum normalized delay shift index, $l_{\text{max}}$ & 1\\
\hline
Maximum normalized Doppler shift index, $k_{\text{max}}$ & 1, 3\\
\hline
Velocity, $v$ & 506.25, 800 km/h\\
\hline
\end{tabular}
\vspace{-3mm}
\end{table}

We first consider generalized OTSM systems having the parameters of Table \ref{table2}. Specifically, the channel coefficients obey $h_i\sim\mathcal{CN}(0,1/P)$. Corresponding to $N=2$ and $N=4$, the maximum speeds are set as $v=506.25$ km/h and $v=800$ km/h, respectively. The maximum normalized delay and Doppler indices are $l_\text{max}=M-1$ and $k_\text{max}=N-1$ \cite{8686339}, where the integer-valued normalized delay and Doppler indices of the $i$th path are considered as $a_i\in\mathcal{U}[1,l_\text{max}]$ ($a_1=0$) and $b_i\in\mathcal{U}[-k_\text{max},k_\text{max}]$, respectively. 

\begin{figure}[htbp]
\vspace{-0.4cm}
\centering
\begin{minipage}[htbp]{\linewidth}
\centering
\includegraphics[width=0.9\linewidth]{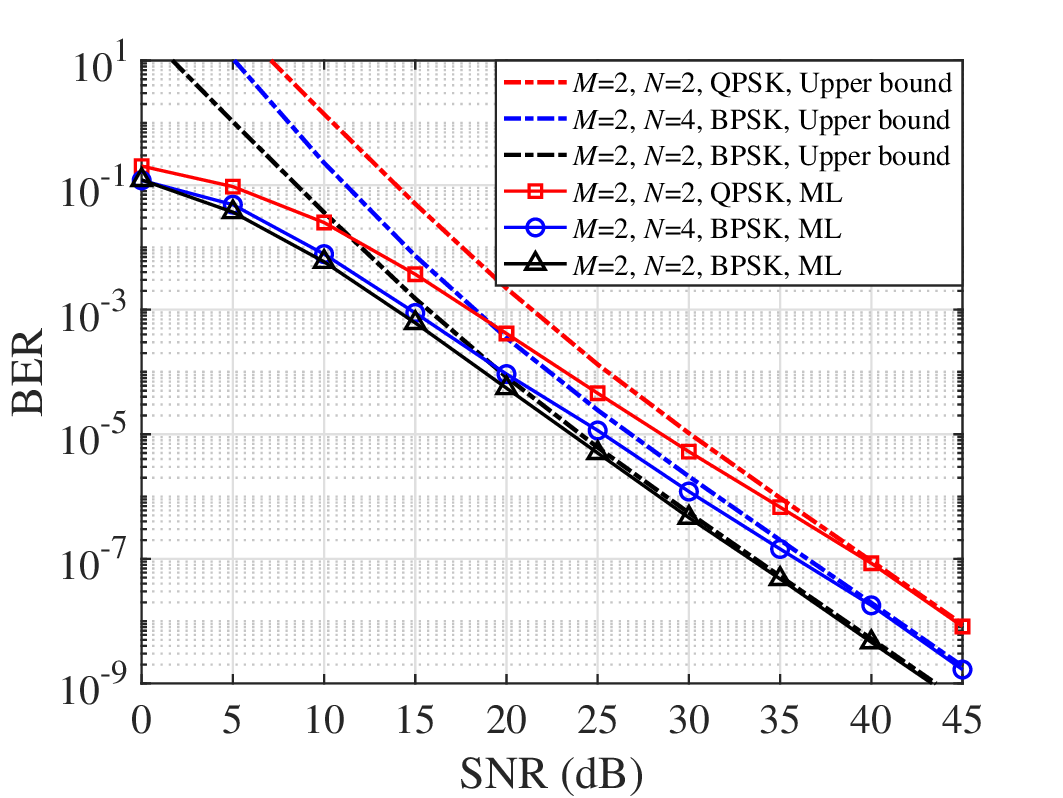}
\vspace{-1em}
\caption{BER performance of the ML detector and the upper bound of generalized OTSM systems with ``$M=2, N=2, \text{QPSK}$", ``$M=2, N=4, \text{BPSK}$" and ``$M=2, N=2, \text{BPSK}$" based on \eqref{Eq42}.}
\label{Figure5}
\end{minipage}
\hspace{0.01in}
\begin{minipage}[htbp]{\linewidth}
\centering
\includegraphics[width=0.9\linewidth]{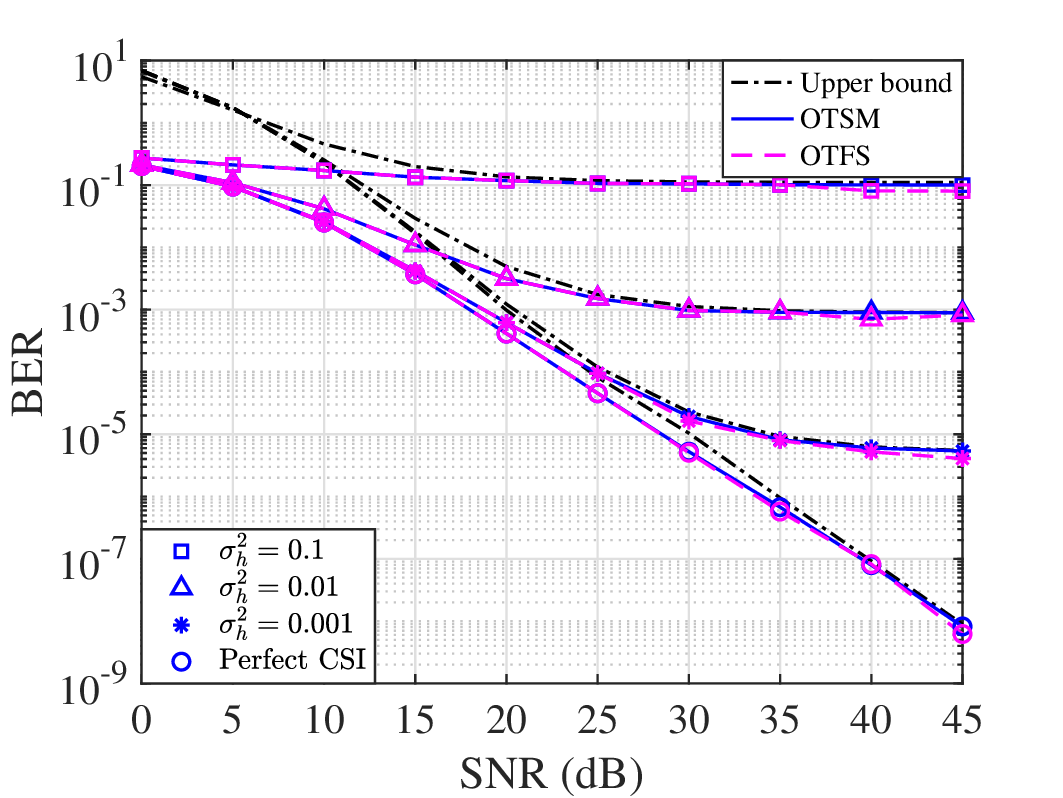}
\vspace{-1em}
\caption{BER performance of both ``$M=2, N=2, \text{QPSK}$" OTSM and OTFS systems using ML detector and the upper bound with different values of $\sigma_h^2$ based on \eqref{Eq41} and \eqref{EqA11}.}
\label{Figure5-1}
\end{minipage}
\end{figure}

In Fig. \ref{Figure5}, the BER performance of the ML detector and the theoretical upper bounds are investigated for parameter settings of ``$M=2, N=2, \text{QPSK}$", ``$M=2, N=4, \text{BPSK}$" and ``$M=2, N=2, \text{BPSK}$". Based on the simulation results of Fig. \ref{Figure5}, we have the following observations. Firstly, it is clear that the upper bound becomes very tight as the SNR values increase. Furthermore, for a given modulation scheme, the OTSM system associated with $MN=4$ an approximately 3 dB lower SNR at BER=$10^{-7}$ than the $MN=8$ OTSM system. This is because the value of $e(\pmb{x}^c,\pmb{x}^e)$ becomes higher upon increasing the $MN$ values, leading a higher value of $P_e$ in \eqref{Eq41}.

Fig. \ref{Figure5-1} further plots the BER performance of both ``$M=2, N=2, \text{QPSK}$" OTFS and OTSM systems using the ML detector and the corresponding upper bounds under imperfect channel estimation conditions. The channel estimation variance is set as $\sigma_h^2=0.1, 0.01$ and $0.001$, respectively. It can be observed from Fig. \ref{Figure5-1} that, as expected, a lower value of $\sigma_h^2$ results in an improved BER performance. Moreover, similar to Fig. \ref{Figure5}, the theoretical upper bounds become tight, as the SNRs escalate. Furthermore, OTSM attains a similar BER performance to that of OTFS under all $\sigma_h^2$ conditions, which is consistent with the simulations in \cite{thaj2021orthogonal}. Finally, we observe that there are BER error floors at $10^{-1}$, $10^{-3}$ and $5\times 10^{-6}$ for $\sigma_h^2=0.1, 0.01$ and $0.001$, respectively. This is because the channel estimation error dominates the BER performance, even if the value of SNR escalates.

Next, we characterize the BER performance and complexity of the proposed VAMP-EM detectors in both uncoded and LDPC-coded systems, the parameters employed are shown in Table \ref{table3}. Explicitly, we consider generating the channel excess tap delay and impulse response power profiles according to the nine-path extended vehicular A (EVA) channel model \cite{thaj2021orthogonal}. The $i$th path normalized Doppler shift index is generated upon the Jake's spectrum, i.e. $k_i=k_\text{max}\cos(\phi_i)$, where $\phi_i\in\mathcal{U}[0,2\pi]$.
\begin{table}[htbp]
\footnotesize
\centering
\caption{Simulation parameters for characterizing the overall performances of detectors}
\label{table3}
\begin{tabular}{l|l}
\hline
\textbf{Parameters} & \textbf{Values} \\
\hline
\hline
Maximum delay-domain grid index, $M$ & 16 \\  
\hline
No. of OTSM symbol, $N$ & 16 \\
\hline
Carrier frequency, $f_c$ & 16 GHz\\
\hline
Subcarrier spacing, $\Delta f$ & 60 kHz\\
\hline
Channel model & EVA channel\\
\hline
No. of path, $P$ & 9\\
\hline
Length of ZP, $L_{\text{ZP}}$ & 4\\
\hline
Velocity, $v$ & 240, 480 km/h\\
\hline
Damping factor, $\theta$ & 0.8\\
\hline
Tolerance parameter, $\epsilon$ & $10^{-10}$\\
\hline
Maximum no. of AMP iterations, $T_{\text{AMP}}$ & 6\\
\hline
Maximum no. of UAMP iterations, $T_{\text{UAMP}}$ & 10\\
\hline
Maximum no. of VAMP-EM iterations, $T$ & 4\\
\hline
Maximum no. of VAMP-EM inner iterations, $T_1$, $T_2$ & 2, 1\\
\hline
Maximum no. of GS iterations, $T_{\text{GS}}$ & 50\\
\hline
LDPC-coded rate, $R$ & $1/2$, $3/4$\\
\hline
\end{tabular}
\end{table}
\begin{figure*}[htbp]
\centering
\vspace{-0.6cm}
\subfigure[]{\label{Figure6-1}\includegraphics[width=0.32\linewidth]{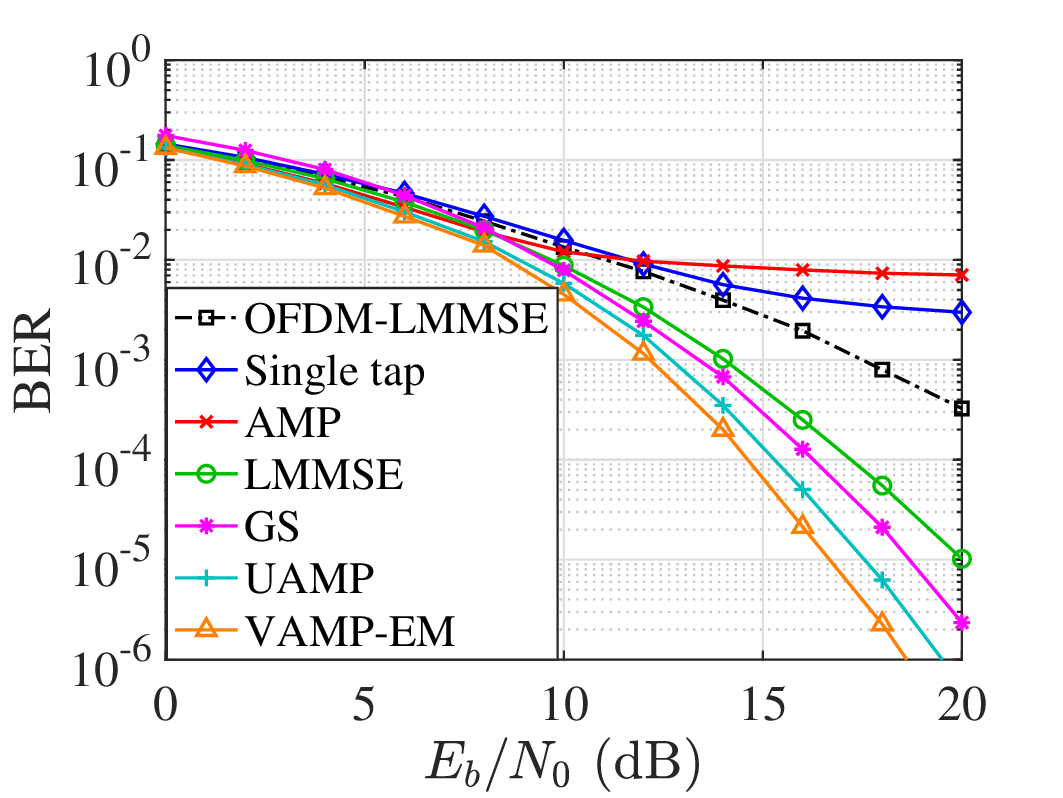}}
\subfigure[]{\label{Figure6-2}\includegraphics[width=0.32\linewidth]{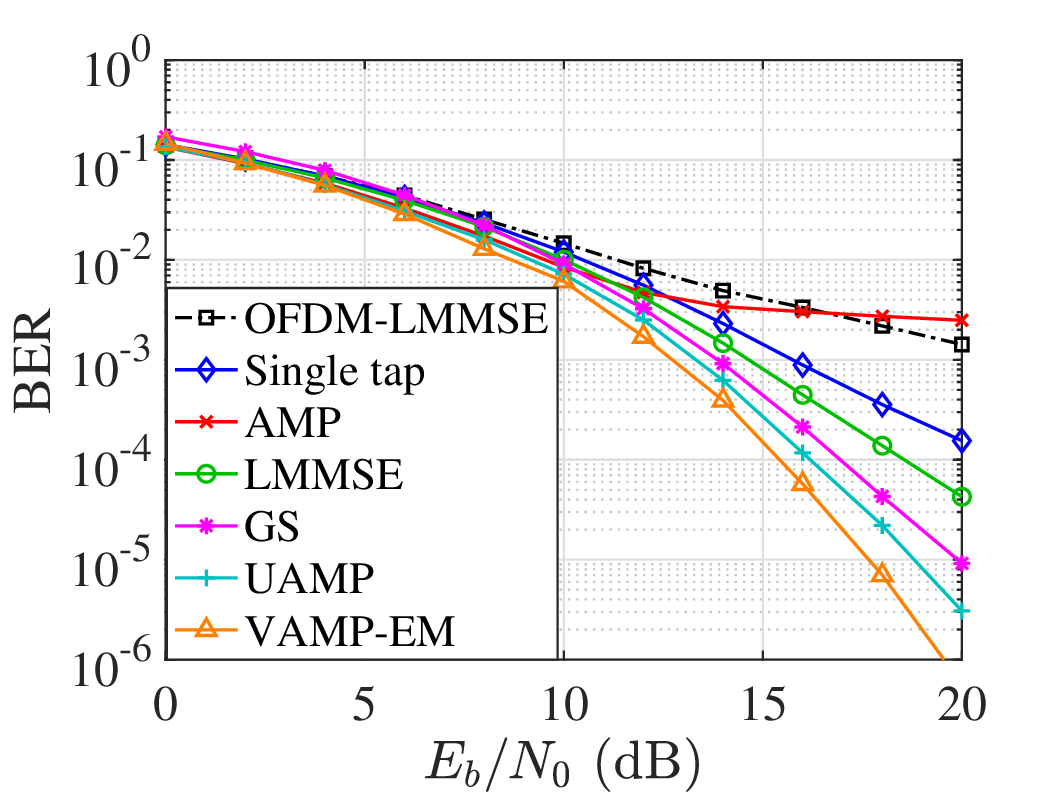}}
\subfigure[]{\label{Figure6-3}\includegraphics[width=0.32\linewidth]{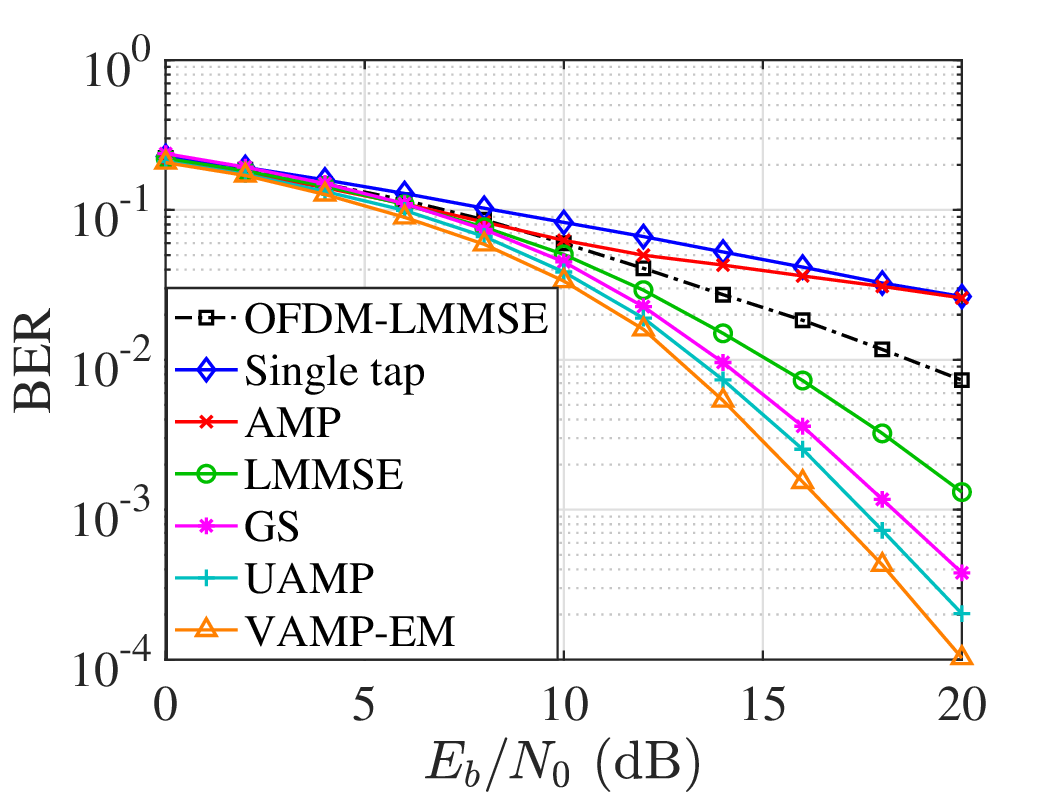}}
\caption{BER performance of the single tap \cite{thaj2021orthogonal}, AMP, LMMSE, GS \cite{thaj2021orthogonal1}, {UAMP \cite{yuan2021iterative}} and the proposed VAMP-EM detection for the OTSM system {with integer-valued delay and Doppler shifts} communicating over time-varying EVA channels operating at \subref{Figure6-1} $(K=4,v=480 km/h)$ \subref{Figure6-2} $(K=4, v=240 km/h)$ \subref{Figure6-3} $(K=16,v=480 km/h)$.}
\vspace{-2em}
\label{Figure6}
\end{figure*}

In Fig. \ref{Figure6}, the BER performance of the single tap detection of \cite{thaj2021orthogonal}, of the AMP, LMMSE, GS \cite{thaj2021orthogonal1}, UAMP \cite{yuan2021iterative} and of the proposed VAMP-EM detectors is investigated for different modulation schemes and relative speeds with integer-valued delay and Doppler shifts. Moreover, the conventional OFDM system associated with LMMSE detection is exploited as a benchmark. We have the following observations based on Fig. \ref{Figure6}. Firstly, we observe that the proposed VAMP-EM detector attains a significantly better BER performance than the other detectors. This is because the \emph{a priori} information of the modulation scheme is exploited in the VAMP-EM, leading to a BER performance gain. In contrast to the AMP detector, the VAMP-EM detector is derived based on the vector-valued factor graph, making VAMP-EM capable of handling ill-conditioned OTSM channel matrices. Secondly, in the SNR region above 12 dB, the AMP detector has a BER floor. This is because the AMP cannot guarantee convergence in the face of practical sensing matrices. Thirdly, observe in Fig. \ref{Figure6} \subref{Figure6-1} and Fig. \ref{Figure6} \subref{Figure6-2} for the BER values of $10^{-5}$, and in Fig. \ref{Figure6} \subref{Figure6-3} for $10^{-3}$, that the VAMP-EM detector requires 2.5 dB lower SNR than the GS detector. At BERs of $10^{-5}$ in Fig. \ref{Figure6} \subref{Figure6-1} and $10^{-3}$ in Fig. \ref{Figure6} \subref{Figure6-3}, our VAMP-EM is capable of attaining 1 dB SNR gain compared to the UAMP detector, while this performance gap escalates to about 1.5 dB at a BER of $10^{-5}$ in Fig. \ref{Figure6} \subref{Figure6-2}. These observations can be explained by the analytical results shown in Section \ref{Section4-4}. Specifically, the VAMP-EM relies on the statistics gleaned from the last iteration and performs LMMSE estimation at every iteration. Moreover, as shown in Fig. \ref{Figure6}, when the relative speed is fixed, the smaller the constellation size, the better the BER performance becomes. This is because a high constellation size exhibits a low minimum Euclidean distance between the symbols. Additionally, it should be noted that all the other counterparts are required to know the value of noise variance. Therefore, the simulation results shown in Fig. \ref{Figure6} indicate that our proposed VAMP-EM detector success in accurately estimating the noise variance $\gamma_n$. Furthermore, the OTSM BER performance of both the single tap and of the AMP detectors is even worse than that of the classic OFDM system employing LMMSE detection at moderate SNR values for $v=480$ km/h and at high SNR values at $v=240$ km/h, respectively. This is because the BER performance of the single tap detector suffers from high ICI, and the AMP detector struggles to provide precise detection results since $\pmb{H}$ is non-Gaussian. Finally, observe from Fig. \ref{Figure6} \subref{Figure6-1} and Fig. \ref{Figure6} \subref{Figure6-2} that the BER performance of detectors does not successively suffer upon increasing the vehicular velocity, which is consistent with the simulation results in \cite{8424569,li2021performance,9772003}.
\begin{figure}[htbp]
\centering
\begin{minipage}[htbp]{\linewidth}
\centering
\vspace{-1mm}
\includegraphics[width=0.9\linewidth]{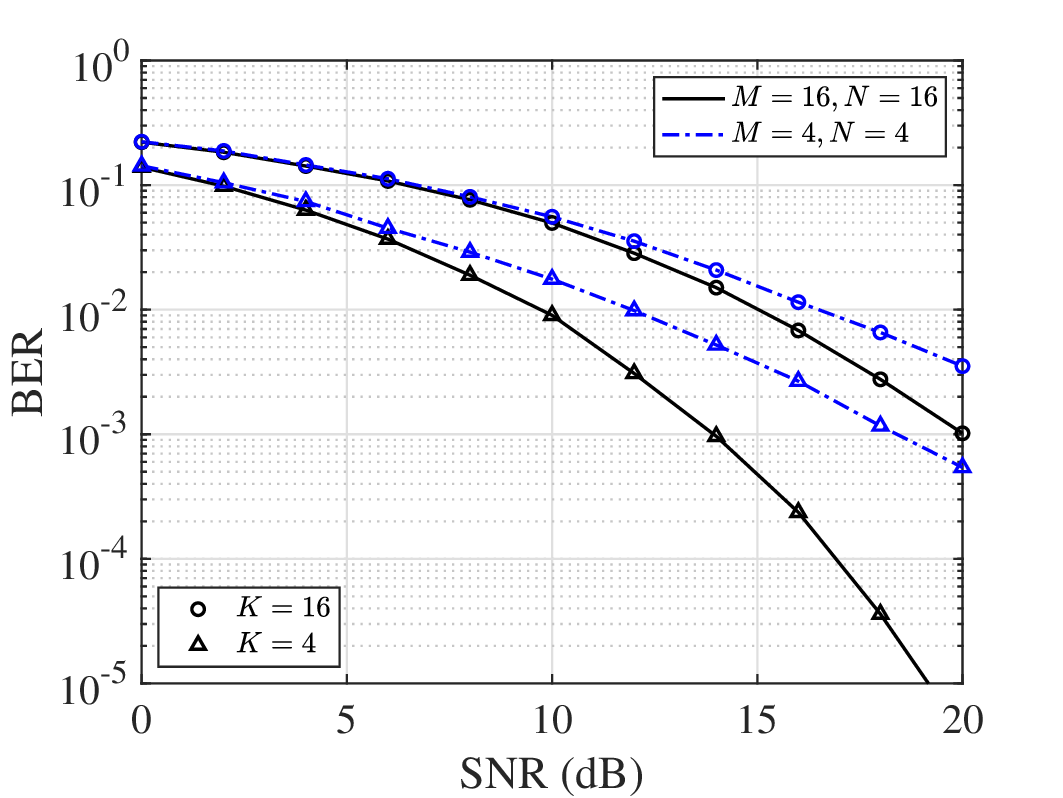}
\vspace{-1em}
\caption{{BER performance of the LMMSE detector with non-integer delay and Doppler shifts operating at different modulation orders as well as different values of $M$ and $N$.}}
\label{Figure7-1}
\end{minipage}
\hspace{0.01in}
\begin{minipage}[htbp]{\linewidth}
\centering
\includegraphics[width=0.9\linewidth]{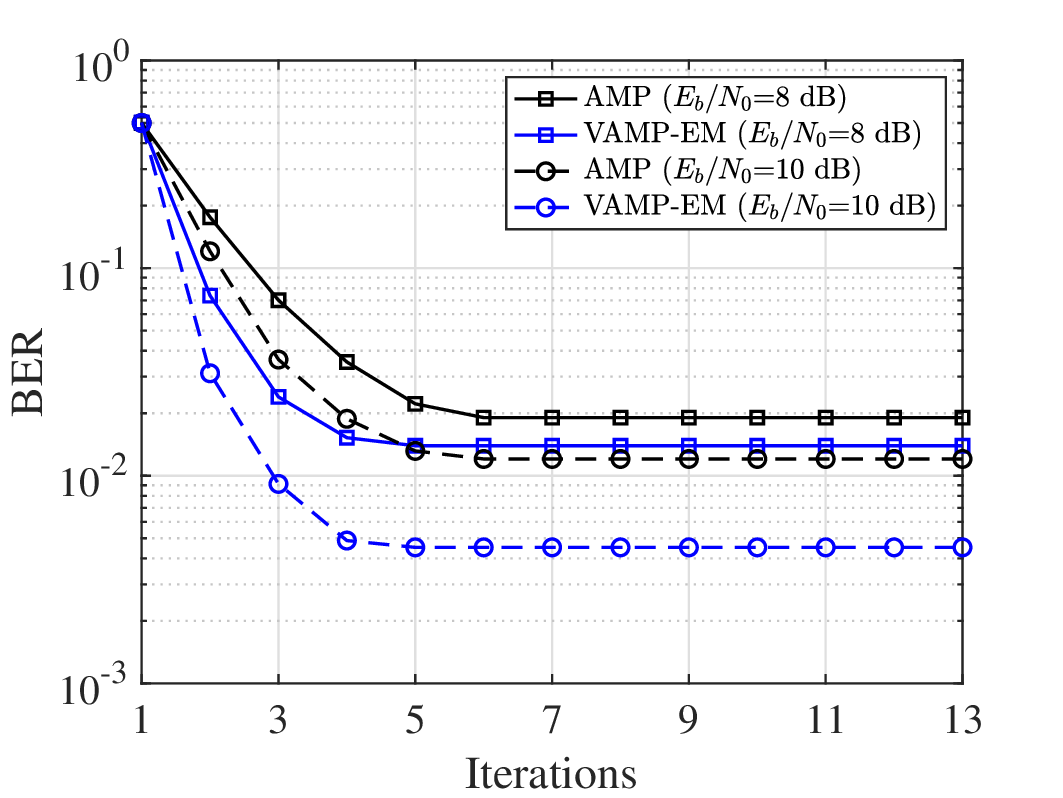}
\vspace{-1em}
\caption{BER performance of the AMP and the proposed VAMP-EM detector versus iteration number operation at $E_b/N_0$=8 dB and $E_b/N_0$=9 dB.}
\label{Figure7}
\end{minipage}
\vspace{-2.0em}
\end{figure}

Furthermore, in Fig. \ref{Figure7-1}, we provide the BER performance of the LMMSE detector for fractional delay and Doppler shifts operating at different numbers of $M$ and $N$, as well as different modulation orders. It is observed from Fig. \ref{Figure7-1} that given the values of $M$ and $N$, a lower modulation order leads to a better BER performance, as expected. More specifically, at a BER of $10^{-3}$, the $(M,N)=(16,16)$ QPSK-modulated system is capable of achieving about 6 dB SNR gain compared to its 16QAM-modulated counterpart. Furthermore, the higher the value of $MN$, the better the BER performance becomes. Explicitly, under the condition of $K=4$, the OTSM system using $(M,N)=(16,16)$ requires about 4.5 dB lower SNR than the $(M,N)=(4,4)$ system at a BER of $10^{-3}$. This can be explained by the fact that a higher value of $MN$ leads to a lower BER, as expected from \eqref{Eq42}.

In Fig. \ref{Figure7}, we investigate the BER performance of both the AMP detector and of our proposed VAMP-EM detector versus the number of detector iterations. The maximum number of iterations are $T_\text{AMP}=T=13$, and the remaining parameters are the same as in Fig. \ref{Figure6} \subref{Figure6-1}. Based on the results, we have the following observations. Firstly, the VAMP-EM detector achieves a better BER than the AMP detector, which is in accord with the observations from Fig. \ref{Figure6}. Moreover, we can see that the proposed VAMP-EM detector converges for $T=4$, yielding a better convergence performance than the AMP detector ($T_\text{AMP}=6$). This is because we incorporate the auto-tuning part into the VAMP-EM detector, which allows the VAMP-EM to recover the transmitted symbols and to estimate the noise variance $1/\gamma_n$ jointly with an improved convergence rate.

\begin{figure}[htbp]
\centering
\begin{minipage}[htbp]{\linewidth}
\centering
\includegraphics[width=0.9\linewidth]{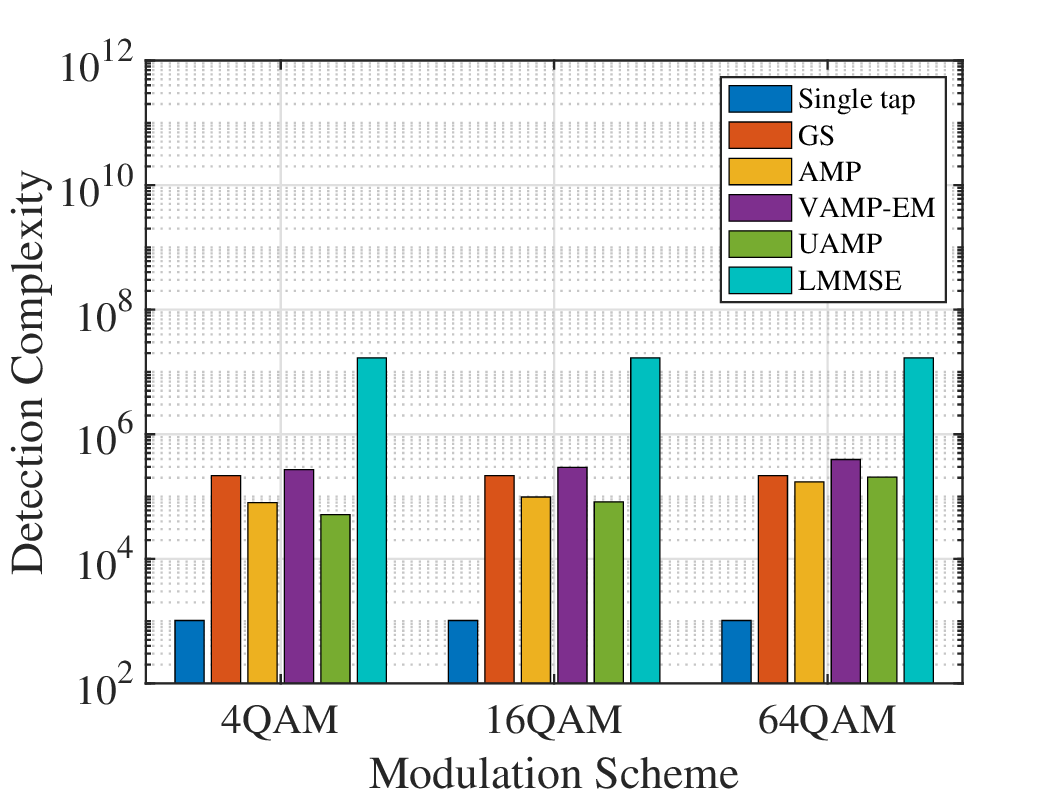}
\vspace{-1em}
\caption{Detection complexity of OTSM using LMMSE, single tap, GS, AMP, UAMP and VAMP-EM detectors.}
\label{Figure8}
\end{minipage}
\hspace{0.01in}
\begin{minipage}[htbp]{\linewidth}
\centering
\includegraphics[width=0.9\linewidth]{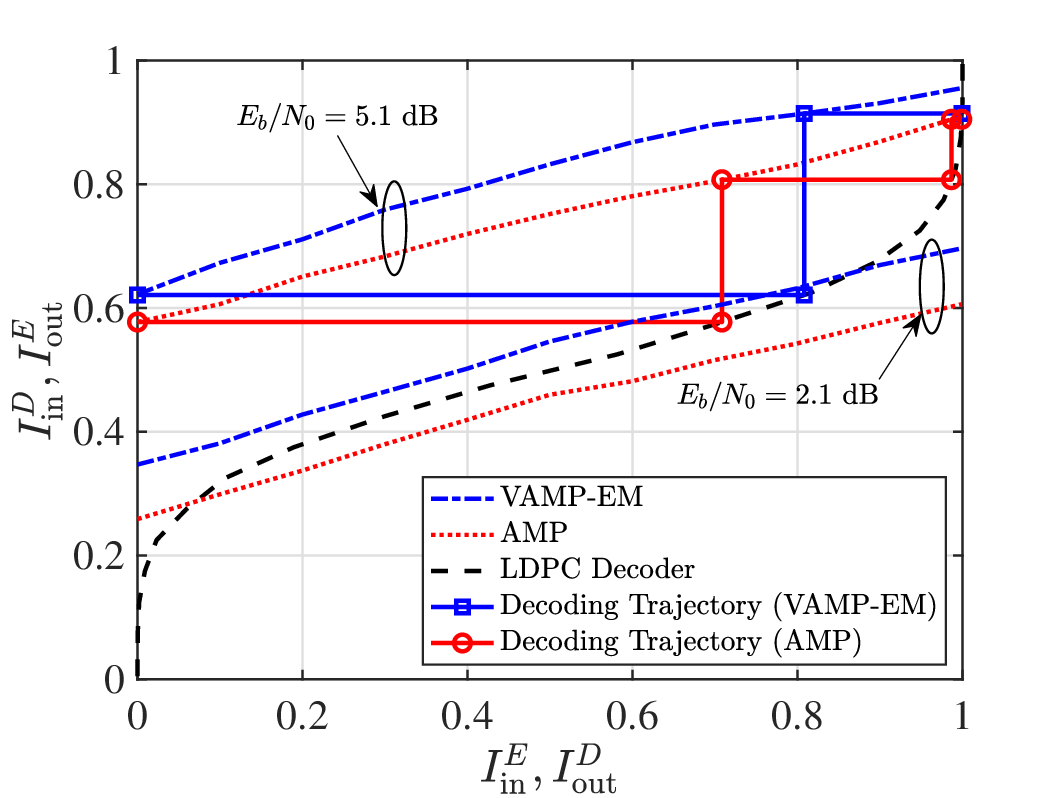}
\vspace{-1em}
\caption{EXIT charts of the proposed VAMP-EM detector and the AMP detector. The decoding trajectories are shown for the 1/2-rate LDPC decoder and for the proposed detector at $E_b/N_0=2.1$ dB and $E_b/N_0=5.1$ dB.}
\label{Figure9}
\end{minipage}
\vspace{-2.0em}
\end{figure}

Fig. \ref{Figure8} shows the detection complexity of LMMSE, VAMP-EM, AMP, GS, UAMP and single tap detectors. The parameters are consistent with those in Fig. \ref{Figure6} \subref{Figure6-1}. It can be seen that the complexity of VAMP-EM is only slightly higher than that of the AMP, GS and UAMP detectors. But they are still of the same order. By contrast, the LMMSE detector imposes the highest complexity. This is because the VAMP-EM detector performs LMMSE and noise variance estimation at every iteration, while the UAMP detector does not involve noise variance estimation under this condition. Based on Fig. \ref{Figure6} and Fig. \ref{Figure8}, it can be observed that the proposed VAMP-EM attains a better BER performance at the cost of a slighter higher complexity than that of the UAMP detector. On the other hand, since the complexity of our VAMP-EM detector is still higher than that of the single-tap detector, we will consider further reducing the VAMP-EM complexity in our future work. Moreover, based on Section \ref{Section4-5}, it can be readily observed that the complexities of the AMP and the VAMP-EM corresponding to large-scale OTSM systems are dominated by the value of $MN^2L$ and $M^2N^2$, respectively. Based on Section \ref{Section2}, it can be readily shown that the throughput of the 64QAM-modulated system is three times higher than that of 4QAM. Hence for the 64QAM OTSM system, the complexities per bit of the AMP-family detectors are nearly three times lower than that of 4QAM. Furthermore, based on Fig. \ref{Figure6}, Fig. \ref{Figure7}, and Fig. \ref{Figure8}, we conclude that the VAMP-EM detector strikes a compelling BER vs. complexity trade-off.

Then, we evaluate the overall performance of the AMP and VAMP-EM turbo receivers in our rate-$R$ LDPC-coded OTSM systems of Fig. \ref{Figure1} and the sum-product decoding algorithm is employed \cite{mackay1997near}. The remaining parameters are consistent with those in Fig. \ref{Figure6} \subref{Figure6-1}. In Fig. \ref{Figure9}, the EXIT curves of the proposed AMP-family-based detectors and that of the rate-1/2 LDPC decoder are investigated, where the superscripts \emph{a} and \emph{e} represent \emph{a priori} and extrinsic information, while the subscripts \emph{det} and \emph{dec} indicate the detector and the decoder, respectively \cite{el2013exit}. As shown in Fig. \ref{Figure9}, the EXIT-tunnel between the detector's and the decoder's curve remain closed at $E_b/N_0$=2.1 dB. By contrast, an open EXIT-tunnel emerges at $E_b/N_0$=5.1 dB, which implies that the proposed AMP and VAMP-EM algorithms are capable of converging in this case. Moreover, in Fig. \ref{Figure9}, the stair-case-shaped decoding trajectories at $E_b/N_0$=5.1 dB between the LDPC decoder and the inner detectors are provided for characterizing the processes of extrinsic information exchanges. It can be observed that the AMP turbo receiver requires at least five iterations between the detector and the LDPC decoder for approaching the maximum mutual information point, while only three corresponding iterations are required by the proposed VAMP-EM turbo receiver to converge around $E_b/N_0$=5.1 dB. This illustrates that the VAMP-EM detector is capable of attaining an improved convergence performance.
\begin{figure}[htbp]
\centering
\begin{minipage}[htbp]{\linewidth}
\centering
\includegraphics[width=0.9\linewidth]{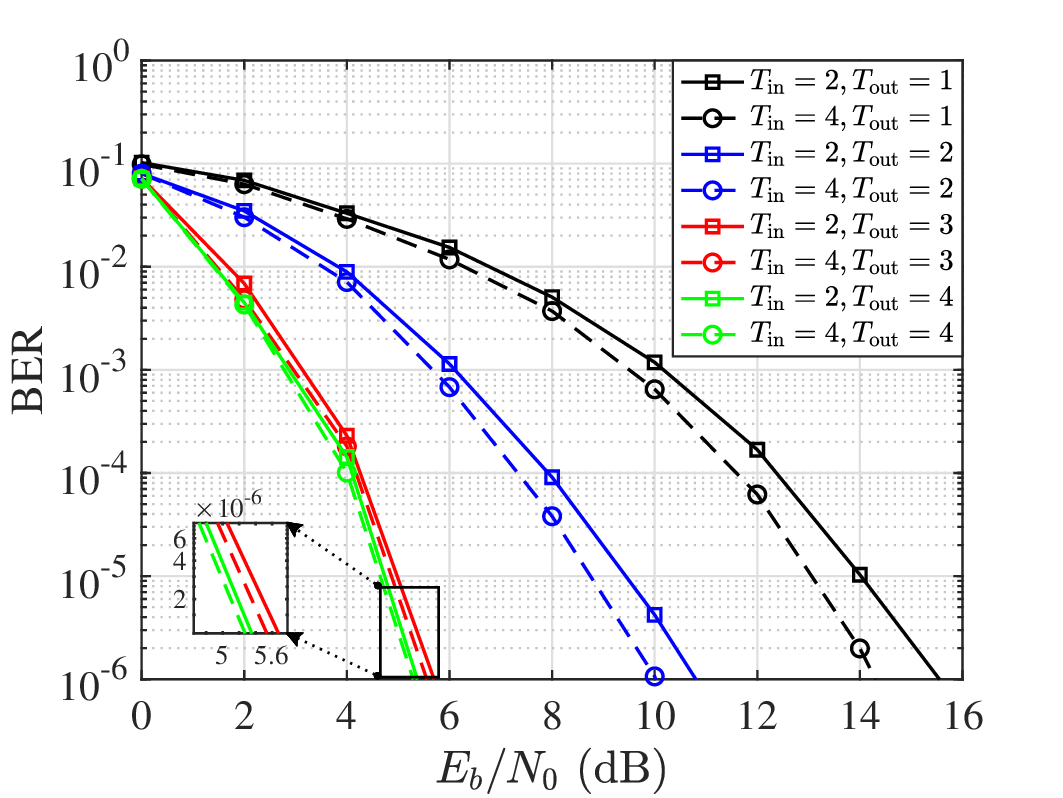}
\vspace{-1em}
\caption{BER performance of the proposed VAMP-EM detector for the rate-1/2 LDPC-coded OTSM system with different numbers of inner decoding iteration and outer turbo detection iteration.}
\label{Figure10}
\end{minipage}
\hspace{0.01in}
\begin{minipage}[htbp]{\linewidth}
\centering
\includegraphics[width=0.9\linewidth]{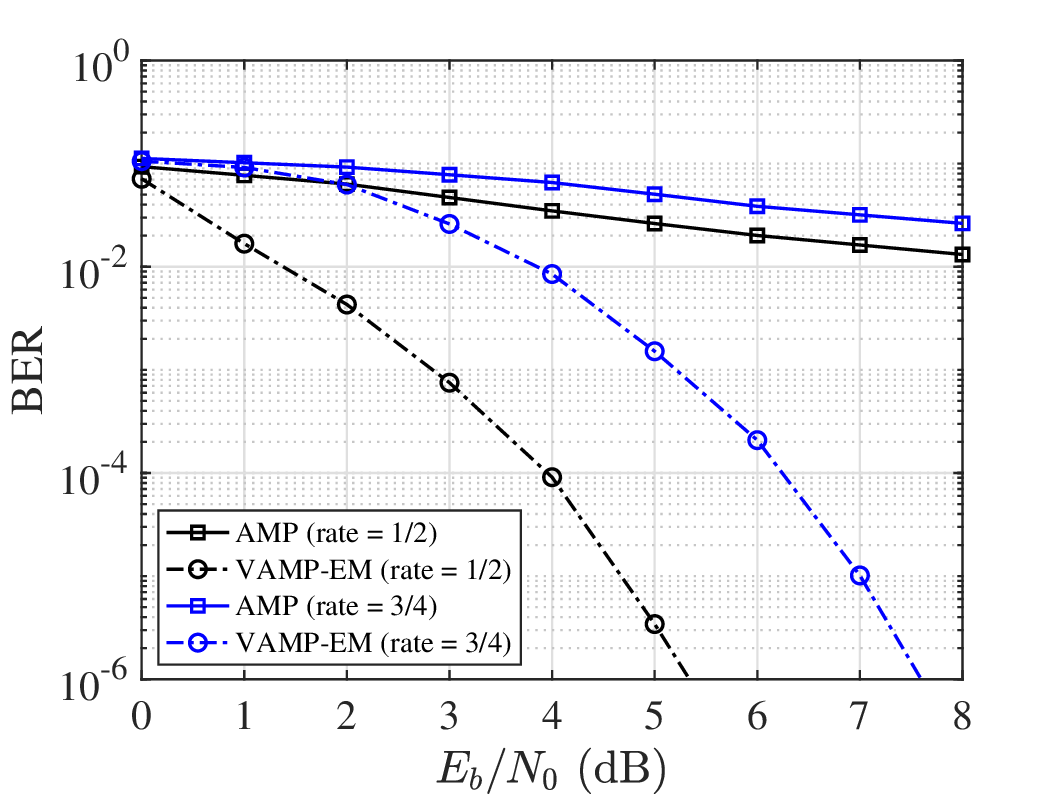}
\vspace{-1em}
\caption{BER performance of the rate-1/2 and rate-3/4 LDPC coded OTSM systems using the AMP detector and the proposed VAMP-EM detector.}
\label{Figure11}
\end{minipage}
\vspace{-1.5em}
\end{figure}

In Fig. \ref{Figure10}, the BER performance of the rate-1/2 LDPC coded OTSM system using the proposed VAMP turbo receivers relying on different numbers of inner decoding iteration $T_{\text{in}}$ and outer detection iteration $T_{\text{out}}$ are evaluated. From Fig. \ref{Figure10}, we have the following observations. Firstly, the BER performance of the VAMP-EM turbo receiver improves as $T_{\text{in}}$ increases. Additionally, the BER performance can be significantly improved by using a higher value of $T_{\text{out}}$. However, the corresponding BER performance gain becomes marginal in the case of $T_{\text{out}}$ being higher than 3, which implies that the combination of the number of inner iterations $T_{\text{in}}$ and outer iterations $T_{\text{out}}$ can be adjusted to strike a performance vs. complexity trade-off. Finally, there is no substantial system performance gain beyond $E_b/N_0$=5.1 dB, and all the above-mentioned observations are consistent with the predictions of the EXIT analysis of Fig.~\ref{Figure9}.

The BER performance of the AMP and VAMP turbo receivers using 1/2-rate and 3/4-rate LDPC coding is investigated in Fig. \ref{Figure11}, where the numbers of inner and external iterations are set as $T_{\text{in}}=4$ and $T_{\text{out}}=4$, respectively. Based on Fig. \ref{Figure6} \subref{Figure6-1} and Fig. \ref{Figure11}, we observe that the BER performance gaps between the AMP and VAMP-EM turbo receivers become larger than that in the uncoded system. Moreover, the lower the LDPC coded rate, the more significant the BER performance gap and the better the BER performance becomes. Furthermore, at a BER of $10^{-6}$, it can be observed that the coding gain is about 2.2 dB associated with $R=1/2$ compared to $R=3/4$. Finally, similar to uncoded systems, due to the ill-conditioning of the DS-domain channel matrix $\pmb{H}$, the convergence performance of the AMP turbo receiver cannot be ensured. Hence, we can observe an error floor at BER=$10^{-2}$.

\section{Conclusion}\label{Section7}
The performance analysis of OTSM systems was provided. Then, based on our theoretical derivations, the input-output relationship of generalized OTSM systems was established. Furthermore, the asymptotical BER upper bound of OTSM has been derived. Our simulation results have shown that the upper bound becomes increasingly tighter upon increasing the SNR. Then, an AMP-aided OTSM detection framework was conceived. By exploiting the statistics of the transmitted symbols as the \emph{a priori} information, a novel VAMP-EM detector was proposed. Explicitly, the EM algorithm was invoked for simultaneously recovering the transmitted symbol and for estimating the noise variance, making the proposed VAMP-EM detector more practical than its conventional counterparts. For further improving the convergence speed attained, an auto-tuning algorithm was conceived as part of the VAMP-EM approach. Our simulation results have demonstrated that the VAMP-EM algorithm is capable of attaining a better convergence and BER performance than the state-of-the-art detection schemes, despite its relatively low complexity. Then both an AMP and a VAMP-EM-based turbo receiver were proposed for LDPC-coded OTSM systems. Simulation results have been offered for characterizing the overall system performance. The VAMP-EM turbo receiver exhibits excellent convergence properties by exploiting the power of iterations between the detector and the decoder. 
\begin{appendices}
\section{Generalized EM framework based on Gibbs sampling}\label{Appendix}
We consider solving the learning $\gamma_n$ problem by leveraging the EM algorithm as illustrated in \eqref{Eq69} and \eqref{Eq70}. Hence, the process of updating $\gamma_n$ can be rewritten as
\begin{align}\label{Eq71}
	V\left(\gamma_n; \hat{\gamma}^{t}_n\right)&=-\mathbb{E}[\ln p(\pmb{x};\gamma_n)|\pmb{y};\hat{\gamma}_n^t]-\mathbb{E}[\ln p(\pmb{y}|\pmb{x};\gamma_n)|\pmb{y};\hat{\gamma}_n^t]\nonumber\\
	&=-\mathbb{E}[\ln p(\pmb{x};\gamma_n)|q^t]-\mathbb{E}[\ln p(\pmb{y}|\pmb{x};\gamma_n)|q^t]\nonumber\\
	&=J(q^t,q^t;\gamma_n)+\text{const},
\end{align}
where $q^t=p(\pmb{x}|\pmb{y};\hat{\gamma}_n^t)$ and the Gibbs free energy is given by $J(q_1,q_2;\gamma_n)\triangleq\ D_\text{KL}[q_1||p(\pmb{y}|\pmb{x};\gamma_n)]+H(q_2)$. The EM processes of \eqref{Eq69} and \eqref{Eq70} can be expressed as \cite{neal1998view} 
\begin{align}\label{Eq73}
q^t=p(\pmb{x},\pmb{y};\gamma_n^t)
	\end{align}
\begin{align}\label{Eq74}
	\hat{\gamma}_n^t=\arg \min _{\gamma_n}J(q^t,q^t;\gamma_n).
\end{align}
It should be noticed that we can simplify $J(q_1,q_2;\gamma_n)$ as \cite{vila2013expectation}
\begin{align}\label{Eq75}
J(q^t,q^t;\gamma_n)&=-\ln p(\pmb{y};\gamma_n)+D_\text{KL}[q^t||p(\pmb{y}|\pmb{x};\gamma_n)]\nonumber\\
&\geq -\ln p(\pmb{y};\gamma_n)
\end{align}	
for any $q^t$ since $D_\text{KL}\geq0$. Therefore, a tighter upper bound of $-\ln p(\pmb{y};\gamma_n)$ is obtained by choosing $q^t$ properly.
\end{appendices}
\vspace{-0.5em}
\renewcommand{\refname}{References}
\mbox{} 
\nocite{*}
\bibliographystyle{IEEEtran}
\bibliography{OTSM.bib}

\begin{thebibliography}{10}

\bibitem{donoho2009message}
David~L Donoho, Arian Maleki, and Andrea Montanari.
\newblock Message-passing algorithms for compressed sensing.
\newblock {\em Proceedings of the National Academy of Sciences},
  106(45):18914--18919, 2009.

\bibitem{el2013exit}
Mohammed El-Hajjar and Lajos Hanzo.
\newblock {EXIT} charts for system design and analysis.
\newblock {\em IEEE Communications Surveys \& Tutorials}, 16(1):127--153, 2013.

\bibitem{DBLP:journals/corr/FletcherSSR17}
Alyson~K. Fletcher, Mojtaba Sahraee{-}Ardakan, Philip Schniter, and Sundeep
  Rangan.
\newblock Rigorous dynamics and consistent estimation in arbitrarily
  conditioned linear systems.
\newblock {\em CoRR}, abs/1706.06054, 2017.

\bibitem{DBLP:journals/corr/abs-1802-02623}
Ronny Hadani and Anton Monk.
\newblock {OTFS:} {A} new generation of modulation addressing the challenges of
  {5G}.
\newblock {\em CoRR}, abs/1802.02623, 2018.

\bibitem{hanzo2005ofdm}
Lajos Hanzo, Byungcho Choi, Thomas Keller, et~al.
\newblock {\em OFDM and MC-CDMA for broadband multi-user communications, WLANs
  and broadcasting}.
\newblock John Wiley \& Sons, 2005.

\bibitem{li2021performance}
Shuangyang Li, Jinhong Yuan, Weijie Yuan, Zhiqiang Wei, Baoming Bai, and
  Derrick Wing~Kwan Ng.
\newblock Performance analysis of coded {OTFS} systems over high-mobility
  channels.
\newblock {\em IEEE Transactions on Wireless Communications}, 20(9):6033--6048,
  2021.

\bibitem{9536449}
Shuangyang Li, Weijie Yuan, Zhiqiang Wei, and Jinhong Yuan.
\newblock Cross domain iterative detection for orthogonal time frequency space
  modulation.
\newblock {\em IEEE Transactions on Wireless Communications}, 21(4):2227--2242,
  2022.

\bibitem{9610105}
Fei Liu, Zhengdao Yuan, Qinghua Guo, Zhongyong Wang, and Peng Sun.
\newblock Multi-block {UAMP}-based detection for {OTFS} with rectangular
  waveform.
\newblock {\em IEEE Wireless Communications Letters}, 11(2):323--327, 2022.

\bibitem{mackay1997near}
David~JC MacKay and Radford~M Neal.
\newblock Near {Shannon} limit performance of low density parity check codes.
\newblock {\em Electronics Letters}, 33(6):457--458, 1997.

\bibitem{neal1998view}
Radford~M Neal and Geoffrey~E Hinton.
\newblock A view of the {EM} algorithm that justifies incremental, sparse, and
  other variants.
\newblock In {\em Learning in graphical models}, pages 355--368. Springer,
  1998.

\bibitem{9285313}
Huiyang Qu, Guanghui Liu, Lei Zhang, Shan Wen, and Muhammad~Ali Imran.
\newblock Low-complexity symbol detection and interference cancellation for
  {OTFS} system.
\newblock {\em IEEE Transactions on Communications}, 69(3):1524--1537, 2021.

\bibitem{rangan2019vector}
Sundeep Rangan, Philip Schniter, and Alyson~K Fletcher.
\newblock Vector approximate message passing.
\newblock {\em IEEE Transactions on Information Theory}, 65(10):6664--6684,
  2019.

\bibitem{8516353}
P.~Raviteja, Yi~Hong, Emanuele Viterbo, and Ezio Biglieri.
\newblock Practical pulse-shaping waveforms for reduced-cyclic-prefix {OTFS}.
\newblock {\em IEEE Transactions on Vehicular Technology}, 68(1):957--961,
  2019.

\bibitem{8424569}
P.~Raviteja, Khoa~T. Phan, Yi~Hong, and Emanuele Viterbo.
\newblock Interference cancellation and iterative detection for orthogonal time
  frequency space modulation.
\newblock {\em IEEE Transactions on Wireless Communications},
  17(10):6501--6515, 2018.

\bibitem{8712432}
Subrata Sarkar, Alyson~K. Fletcher, Sundeep Rangan, and Philip Schniter.
\newblock Bilinear recovery using adaptive {Vector-AMP}.
\newblock {\em IEEE Transactions on Signal Processing}, 67(13):3383--3396,
  2019.

\bibitem{9590508}
Suraj Srivastava, Rahul~Kumar Singh, Aditya~K. Jagannatham, and Lajos Hanzo.
\newblock Bayesian learning aided simultaneous row and group sparse channel
  estimation in orthogonal time frequency space modulated {MIMO} systems.
\newblock {\em IEEE Transactions on Communications}, 70(1):635--648, 2022.

\bibitem{steele1999mobile}
Raymond Steele and Lajos Hanzo.
\newblock {\em Mobile radio communications: Second and third generation
  cellular and WATM systems: 2nd}.
\newblock IEEE Press-John Wiley, 1999.

\bibitem{9507331}
Zeping Sui, Shefeng Yan, Hongming Zhang, Lie-Liang Yang, and Lajos Hanzo.
\newblock Approximate message passing algorithms for low complexity {OFDM-IM}
  detection.
\newblock {\em IEEE Transactions on Vehicular Technology}, 70(9):9607--9612,
  2021.

\bibitem{8686339}
GD~Surabhi, Rose~Mary Augustine, and A~Chockalingam.
\newblock On the diversity of uncoded {OTFS} modulation in doubly-dispersive
  channels.
\newblock {\em IEEE Transactions on Wireless Communications}, 18(6):3049--3063,
  2019.

\bibitem{tarokh1998space}
Vahid Tarokh, Nambi Seshadri, and A~Robert Calderbank.
\newblock Space-time codes for high data rate wireless communication:
  Performance criterion and code construction.
\newblock {\em IEEE Transactions on Information Theory}, 44(2):744--765, 1998.

\bibitem{thaj2021orthogonal1}
Tharaj Thaj and Emanuele Viterbo.
\newblock Orthogonal time sequency multiplexing modulation.
\newblock In {\em 2021 IEEE Wireless Communications and Networking Conference
  (WCNC)}, pages 1--7. IEEE, 2021.

\bibitem{9772003}
Tharaj Thaj and Emanuele Viterbo.
\newblock Unitary-precoded single-carrier waveforms for high mobility:
  Detection and channel estimation.
\newblock In {\em 2022 IEEE Wireless Communications and Networking Conference
  (WCNC)}, pages 962--967, 2022.

\bibitem{thaj2021orthogonal}
Tharaj Thaj, Emanuele Viterbo, and Yi~Hong.
\newblock Orthogonal time sequency multiplexing modulation: Analysis and
  low-complexity receiver design.
\newblock {\em IEEE Transactions on Wireless Communications},
  20(12):7842--7855, 2021.

\bibitem{tuchler2002turbo}
Michael Tuchler, Ralf Koetter, and Andrew~C Singer.
\newblock Turbo equalization: principles and new results.
\newblock {\em IEEE Transactions on Communications}, 50(5):754--767, 2002.

\bibitem{van2000ubiquitous}
Charles~F Van~Loan.
\newblock The ubiquitous kronecker product.
\newblock {\em Journal of Computational and Applied Mathematics},
  123(1-2):85--100, 2000.

\bibitem{vila2013expectation}
Jeremy~P Vila and Philip Schniter.
\newblock Expectation-maximization gaussian-mixture approximate message
  passing.
\newblock {\em IEEE Transactions on Signal Processing}, 61(19):4658--4672,
  2013.

\bibitem{9508932}
Zhiqiang Wei, Weijie Yuan, Shuangyang Li, Jinhong Yuan, Ganesh Bharatula, Ronny
  Hadani, and Lajos Hanzo.
\newblock Orthogonal time-frequency space modulation: A promising
  next-generation waveform.
\newblock {\em IEEE Wireless Communications}, 28(4):136--144, 2021.

\bibitem{yuan2021iterative}
Zhengdao Yuan, Fei Liu, Weijie Yuan, Qinghua Guo, Zhongyong Wang, and Jinhong
  Yuan.
\newblock Iterative detection for orthogonal time frequency space modulation
  with unitary approximate message passing.
\newblock {\em IEEE Transactions on Wireless Communications}, 2021.

\bibitem{8933099}
Hongming Zhang, Chunxiao Jiang, Jingjing Wang, Li~Wang, Yong Ren, and Lajos
  Hanzo.
\newblock Multicast beamforming optimization in cloud-based heterogeneous
  terrestrial and satellite networks.
\newblock {\em IEEE Transactions on Vehicular Technology}, 69(2):1766--1776,
  2020.

\bibitem{zhang2018linear}
Hongming Zhang, Chunxiao Jiang, Lie-Liang Yang, Ertugrul Basar, and Lajos
  Hanzo.
\newblock Linear precoded index modulation.
\newblock {\em IEEE Transactions on Communications}, 67(1):350--363, 2018.

\bibitem{zhang2016compressed}
Hongming Zhang, Lie-Liang Yang, and Lajos Hanzo.
\newblock Compressed sensing improves the performance of subcarrier
  index-modulation-assisted {OFDM}.
\newblock {\em IEEE Access}, 4:7859--7873, 2016.

\end{thebibliography}

\end{document}